\documentclass[12pt]{article}

\pdfoutput=1

\usepackage{amsmath,amssymb,amsfonts,amscd,mathrsfs}
\usepackage{xcolor}
\definecolor{darkblue}{rgb}{0.1,0.1,.7}
\usepackage[colorlinks, linkcolor=darkblue, citecolor=darkblue, urlcolor=darkblue, linktocpage]{hyperref} 
\usepackage[square, comma, compress,numbers]{natbib}
\usepackage[]{version}
\usepackage[]{graphicx}
\usepackage[]{latexsym}
\usepackage{geometry}
\usepackage[all,cmtip]{xy}
\usepackage[margin=10pt,font=small,labelfont=bf]{caption}
\usepackage{ifthen}
\usepackage{tikz}
\usepackage{array,setspace,mathrsfs,amsfonts,yfonts,dsfont,bbm,colonequals}

\usepackage{titlesec}
\titleformat*{\section}{\large\bfseries}
\titleformat*{\subsection}{\normalsize\bfseries}
\titleformat*{\subsubsection}{\normalsize\bfseries}
\titleformat*{\paragraph}{\normalsize\bfseries}
\titleformat*{\subparagraph}{\normalsize\bfseries}

\def\be{\begin{equation}}
\def\ee{\end{equation}}
\def\bea{\begin{eqnarray}}
\def\eea{\end{eqnarray}}

\def\cal{\mathcal}

\newcommand\fverb{\setbox\fverbbox=\hbox\bgroup\verb}
\newcommand\fverbdo{\egroup\medskip\noindent%
			\fbox{\unhbox\fverbbox}\ }
\newcommand\fverbit{\egroup\item[\fbox{\unhbox\fverbbox}]}
\newbox\fverbbox

\def\bea{\begin{eqnarray}}
\def\eea{\end{eqnarray}}
\def\be{\begin{equation}}
\def\ee{\end{equation}}

\def\D{\Delta}

\def\la{\lambda}

\def\OO{\mathcal{O}}

\def\DD{\mathcal{D}}

\def \del{\partial}

\newcommand{\du}[2]{_{#1}^{\phantom{#1}#2}} 

\newcommand\restr[2]{{
  \left.\kern-\nulldelimiterspace 
  #1 
  \vphantom{\big|} 
  \right|_{#2} 
  }}

\def\Sd{\mathrm{S}_d}

\def\pd{\partial}

\usepackage{accents}
\newlength{\dhatheight}

\usepackage{dsfont} 

\def\bsub{\begin{subequations}}
\def\esub{\end{subequations}}
\newcommand{\braket}[3]{\langle #1|#2|#3 \rangle}

\newcommand{\ket}[1]{|#1\rangle}

\newcommand{\expec}[1]{\langle #1 \rangle}
\newcommand{\no}[1]{:\! #1 \!:}

\newcommand{\mbb}[1]{{\mathbb #1}}

\newcommand{\mca}[1]{{\mathcal #1}}

\newcommand{\mrm}[1]{{\mathrm #1}}

\def\pd{\partial}

\def\la{\lambda}
\def\DD{\Delta}
\def\Oo{\mathcal{O}}

\def\eps{\epsilon}

\def\tr{\, \mathrm{tr} \,}

\def\unit{\mathds{1}} 

\newcommand{\reef}[1]{(\ref{#1})}
\def\vareps{\varepsilon}
\def\eps{\epsilon}
\newcommand{\beq}{\begin{equation}} 
\newcommand{\eeq}{\end{equation}}
\def\del {\partial} 
 
\def\bZ {\mathbb{Z}}

\def\calO {{\cal O}} 
 
\def\calD {{\cal D}}

\def\calM {{\cal M}}

\def\calA {{\cal A}} 
\def\calB {{\cal B}}

\def\bZ {\mathbb{Z}} 
 
\def\half{{\textstyle\frac 12}}

\def\ge{\geqslant}
\def\le{\leqslant}
\def\geq{\geqslant}
\def\leq{\leqslant}
\newcommand{\diffop}[2]{\ifthenelse{\equal{#2}{1}}{\frac{\mrm{d}}{\mrm{d} #1}}{\frac{\mrm{d}^#2}{\mrm{d} #1^#2}}}

\def\Sd{{\rm S}_d}

\newcommand{\NO}[1]{{:\!#1\!:}}

\numberwithin{equation}{section}
\begin{document}

\vspace*{-.6in} \thispagestyle{empty}
\begin{flushright}
CERN TH/2015-282\\
YITP-SB-15-44\\
DCPT-15/65
\end{flushright}
\vspace{1cm} {\Large
\begin{center}
{\bf Unitarity violation at the Wilson-Fisher fixed point\\ in $4-\eps$ dimensions}\\
\end{center}}
\vspace{1cm}
\begin{center}
{\bf Matthijs Hogervorst$^{a,b,c}$, Slava Rychkov$^{c,b,d}$, Balt C.~van Rees$^{c,e}$ }\\[2cm] 
{
$^{a}$ C.N.\@ Yang Institute for Theoretical Physics, Stony Brook University, USA\\
$^{b}$ Laboratoire de Physique Th\'{e}orique de l'\'{E}cole normale sup\'{e}rieure (LPTENS), Paris, France\\
$^{c}$ CERN, Theory Department, Geneva, Switzerland\\
$^{d}$ Sorbonne Universit\'es, UPMC Univ Paris 06, Facult\'e de Physique, Paris, France\\
$^{e}$ Centre for Particle Theory \& Department of Mathematical Sciences,\\
Durham University, Durham, UK
}
\\
\vspace{1cm}
\end{center}

\vspace{4mm}

\begin{abstract}
We consider the continuation of free and interacting scalar field theory to non-integer spacetime dimension $d$. We find that the correlation functions in these theories are necessarily incompatible with unitarity (or with reflection positivity in Euclidean signature). In particular, the theories contain negative norm states unless $d$ is a positive integer. These negative norm states can be obtained via the OPE from simple positive norm operators,
and are therefore an integral part of the theory. At the Wilson-Fisher fixed point the non-unitarity leads to the existence of complex anomalous dimensions. We demonstrate that they appear already at leading order in the epsilon expansion.
\end{abstract}
\vspace{.2in}
\vspace{.3in}
\hspace{0.7cm}November 2015

\newpage

{
\setlength{\parskip}{0.05in}
\tableofcontents
}
\newpage

\setlength{\parskip}{0.05in}

\section{Introduction}
\label{sec:intro}

Unitarity, or its absence, is a key structural property of any quantum field theory (QFT). 
The purpose of this paper is to present a novel mechanism by which unitarity can be violated for QFTs defined in \emph{non-integer} spacetime dimensions. This mechanism was first pointed out in our recent work~\cite{tcsa}, and here it will be explained in detail. The mechanism is general, but here it will be demonstrated using the example of the scalar $\phi^4$ theory in $d$ dimensions, described by the Lagrangian
\beq
\mca{L} = \frac{1}{2} (\pd \phi)^2 + \frac{1}{4!} g \mu^{4-d}  \phi^4\,. \label{eq:WFlagrangian}
\eeq
As is well known, in $d = 4 - \eps$ dimensions, this theory has a conformal IR fixed point, called the Wilson-Fisher (WF) fixed point~\cite{Wilson:1971dc}. The fixed point coupling $g=g_*$ is given at the lowest order by
\beq g_*/ (4\pi)^2= \eps/3 + O(\eps^2)\,. \label{eq:gcritical} \eeq
Unitarity is violated, we believe, for all non-integer $d$ in the usually considered range ${2\le d \le 4}$.\footnote{\label{note:Miguel}
Continuation to a larger range $1\le d\le 4$ is also sometimes considered (see \cite{Golden:2014oqa} for a recent discussion). The theories in the range $1<d<2$ are also expected to be non-unitary. For the free scalar theory this is obvious, as the dimension of $\phi$ is negative and violates the scalar primary unitarity bound $\max(0, (d-2)/2)$.
} However, for most of the paper we will focus on $d=4-\eps$, $\eps\ll 1$. On the one hand, this will be enough to illustrate the general mechanism. On the other hand, since the theory in this range is weakly coupled, it will be possible to buttress general arguments by explicit perturbative calculations.

The mechanism responsible for these unitarity violations makes use of operators that are present in the spectrum for all non-integer $d$, but that decouple for certain integer $d$. Such operators are known as \emph{evanescent operators} in the QFT literature~\cite{Collins:1984xc}. In standard QFT computations, one is interested in computing observables at integer $d$, typically ${d=4}$. After regulating the theory using dimensional regularization and passing to ${d = 4 - \eps}$ dimensions, such evanescent operators can appear, as has been shown in theories with fermions~\cite{Dugan:1990df,Gracey:2008mf}.\footnote{Evanescent operators are also necessary to renormalize theories whose action involves the epsilon tensor, like the WZNW model \cite{Bos:1987fb}.} Although these operators themselves decouple in the limit ${\eps \to 0}$, they leave an imprint on 4d observables. A well-known example of this phenomenon is furnished by the QCD NLO anomalous dimensions of four-Fermi operators, responsible for the hadronic weak decays~\cite{Buras:1998raa}.

In contrast to all the works just cited, here we will not take the limit $\eps\to0$. This is because we are primarily interested in the WF fixed point, which becomes trivial in this limit. Rather we will keep $\eps$ small but finite. Our main point will be then to show that some of the evanescent operators give rise to states that have negative norm, implying that the $\phi^4$ theory in non-integer dimensions, and a fortiori the WF fixed point, is not unitary. In passing, the present work shows that evanescent operators already occur in theories with a single scalar field, whereas previous QFT computations only encountered such operators when non-scalar fields were present.

There is a simple reason that evanescent operators in the WF fixed point have not been noticed up to now.
This is because most computations have focused on low-dimension operators such as $\phi$, $\phi^2$ and $\phi^4$, which are obviously present in $d=4$. As we will see, to discover the evanescent operators in the scalar theory and to demonstrate their various properties, one has to go pretty high up in operator dimension.\footnote{This is different from the fermionic case where already at the four-Fermi level one finds an infinite number of evanescent operators.} For instance, in this paper we will work with all scalar operators below the cutoff $\Delta_{\max}=23$ in operator dimension. The number of operators grows exponentially fast, and as a result there are thousands of operators below this cutoff. 

To our knowledge, such a systematic investigation of the Hilbert space and spectrum of WF was taken only once before us, by Kehrein et al~\cite{Kehrein:1992fn,Kehrein:1994ff,Kehrein:1995ia}. These papers used a different methodology that did not take evanescent operators into account. As a result, they did not notice the presence of negative-norm states in $4-\eps$ dimensions. In other aspects there are some similarities between our works, and a more detailed comparison will be given in section \ref{sec:Kehrein}. 

The remainder of this paper is organized as follows. Section \ref{sec:freetheory} discusses evanescent operators in the free scalar theory. We will show that their existence implies that the theory in non-integer dimensions has negative-norm states, violating unitary. In addition we mention how these operators fit into representations of the conformal algebra, and we also give a systematic counting of all the evanescent operators around $d = 4$ up to $\Delta_{\max} = 23$.

From section \ref{sec:WF} onwards we consider the interacting theory at the WF fixed point in $4 - \epsilon$ dimensions. We begin with a review of the computation of one-loop anomalous dimensions in section \ref{sec:WF}. In section $\,\,\,\,$\ref{sec:nonunitWF} we discuss the anomalous dimensions of the evanescent operators and show that there is no reason for them to be real-valued. We substantiate this claim with an explicit computation in subsection \ref{sec:complex}: there exist four scalar operators with $\Delta = 23$ (in $d=4$) that get a complex anomalous dimension at one loop. We also briefly explain how the evanescent operators can affect the general computation of higher-loop anomalous dimensions. The implications of our results for the structure of four-point functions are discussed in appendix \ref{app:implications}.

Our final section \ref{sec:finalcomments} offers some general comments, including an overview of the expected structure of the WF fixed point for any $2 \leq d \leq 4$ that stems from our results. In subsection \ref{subsec:bootstrap} we consider the effect of evanescent operators on numerical studies of the crossing symmetry equations.

\section{The free boson in non-integer $d$}
\label{sec:freetheory}
As a first order of business we will have to make the definition of a quantum field theory in non-integer dimensions more precise. We will investigate local operators and their correlation functions, and leave the study of other observables like the S-matrix or nonlocal operators to future work.\footnote{Results on nonlocal operators in the epsilon expansion can be found, for example, in \cite{Gaiotto:2013nva,Chester:2015wao}.} In this section we will consider the free massless scalar field $\phi$ in non-integer dimensions in more detail. Unless otherwise mentioned, we will always work in Euclidean signature. 

For any \emph{integer} $d > 2$, the set of local operators is generated by taking symmetrized normal ordered products, or `words', of the elementary `letters' which consist of zero or more derivatives acting on the fundamental field $\phi$. Each operator therefore looks like
\be
:\del^{n_1} \phi \, \del^{n_2} \phi \ldots \, \del^{n_k} \phi: (x)\,, \label{eq:genericOp}
\ee
where $\del^{n}$ stand for various derivatives of order $n$, some of whose indices may be contracted (see below). Correlation functions and OPEs of these operators can be computed by Wick's theorem.

In this paper we will mainly consider scalar, parity-even operators.\footnote{See however subsection \ref{sec:tensops} for a brief discussion of general tensorial operators, the epsilon tensor and the parity operation.} A basis for such operators is produced by taking operators of the form~\reef{eq:genericOp} and contracting all the indices on the derivatives with the inverse metric $\delta^{\mu \nu}$.
Contracting two derivatives acting on the same $\phi$ is considered to give rise to a zero operator, and such operators are not included in the basis. This is justified, because operators proportional to the equation of motion $\Box \phi = 0$, $\Box:=\del^\mu\del_\mu$, have non-vanishing correlation functions only at coincident points. If one is interested in correlation functions at non-coincident points, as is the case here, one can drop such operators from the start. In the above discussion, this means that we will eliminate operators of the form $\del^n \Box\phi\times(\text{anything})$.

Through the operator-state correspondence the set of operators also defines the Hilbert space of the theory on the $(d-1)$--sphere $S^{d-1}$. As is familiar, the states in this Hilbert space can be decomposed into highest-weight representations of the $d$-dimensional conformal algebra $SO(d+1,1)$ built on top of conformal primary operators satisfying $[K_\mu,\calO(0)]=0$. For our purposes, we will not need to explicitly organize all states in such representations. With some abuse of language, in this work we will sometimes talk about states and operators interchangeably.

Moving to \emph{non-integer} $d$, most of the above story extends in a fairly standard and natural manner. The scaling dimension of $\phi$ is analytically continued from integer $d$, i.e.~${\Delta_\phi = (d-2)/2}$. One does not dwell too much on the range of the index $\mu$ for non-integer $d$.
One also imposes that $\pd_\mu x^\nu = \delta_{\mu}^{\nu}$ and furthermore that the metric satisfies $\delta^{\mu \nu}\delta_{\mu \nu} = d$. This procedure completely defines correlation functions of local operators in flat space, with the exception of those involving the epsilon tensor, which we discuss in more detail below. It basically amounts to the analytic continuation of coefficients multiplying the tensor structures.

We will also need to integrate correlation functions over either all of $\mathbb R^d$ or on $S^{d-1}.$ 
We will use the standard dimensional regularization integration rules \cite{Wilson:1972cf,Collins:1984xc}. In particular the volume of a unit sphere in $\mathbb R^d$ is $\Sd := \text{Vol}(S^{d-1}) = 2 \pi^{d/2}/ \Gamma(d/2)$. As explained in \cite{Wilson:1972cf,Collins:1984xc}, such $d$-dimensional integrals can be properly defined if one considers vectors to have infinitely many components in non-integer dimensions; one may show that they truncate to finite-dimensional integrals when $d$ becomes integer.

Finally, it will be helpful to have a definition of correlation functions on the cylinder $\mathbb R \times S^{d-1}$. We define these formally through the standard Weyl transformation rules from flat space. Mapping to the cylinder is morally equivalent to the radial quantization of the theory, which will be extensively used in the following sections. Notice that if we can integrate correlation functions then we can also define non-local operators, like the generators of the conformal algebra, hence conformal transformations on local operators are defined unambiguously.
We will therefore also be able to group operators into conformal multiplets and distinguish between primary and descendant operators.

Although the above setup defines all local observables needed in this work, some aspects of the continuation to non-integer $d$ remain somewhat mysterious. These issues relate to the representation theory of the algebras $so(d)$ and $so(d+1,1)$ in non-integer $d$, which does not appear to be developed in the mathematical literature.\footnote{
An analytic continuation of the algebra $sl(d)$ to general $d$ has been constructed in Ref.~\cite{feigin1988}.
} One concrete relevant question concerns the meaning of the analytic continuation of the partition function $Z$ of the free boson on $S^1 \times S^{d-1}$ \cite{Cardy:1991kr}. For integer $d$ the partition function counts all states on the sphere with weight one (times $q^\Delta$, where $\Delta$ is the dimension and $q$ is the modular parameter). In analytically continued $Z$, the vector $\ket{\pd_\mu \phi}$ is then counted with weight $d$, whereas we expect that this representation is infinite-dimensional for non-integer $d$. This leads to the conclusion that in non-integer $d$, the partition function $Z$ only counts degrees of freedom in a regulated sense. Is there an intrinsic representation-theoretic meaning to the dimensions of $so(d)$ representations analytically continued to non-integer $d$?
It would be interesting to better understand this question.\footnote{Similar issues arise for $O(N)$ models and $Q$-state Potts models with non-integer $N$ or $Q$, relevant for percolation and self-avoiding walks. There is a wide statistical physics literature on the partition function of such models where the coefficients are polynomials of $N$ or $\sqrt{Q}$, interpreted as quantum dimensions of quantum group representations. We just give two references \cite{francesco_relations_1987,Cardy:2006fg} as points of entry. We thank Jesper Jacobsen and Hirohiko Shimada for the discussions.
}

\subsection{Evanescent operators}
\label{sec:evan}
Consider the sequence of scalar, parity-even operators of the form 
\beq
\label{eq:Zn} R_n(x) \colonequals \delta^{\mu_1[\nu_1}\delta^{\vert\mu_2\vert\nu_2} \dots\delta^{\vert\mu_n\vert\nu_n]}   \no{ M_{\mu_1 \nu_1}(x) \dotsm M_{\mu_n \nu_n}(x)} \,,
\eeq
where $n=1,2,\ldots$ is a positive integer and we have introduced the shorthand notation
\beq 
M_{\mu \nu}(x) = \pd_\mu \pd_\nu \phi(x)\,. 
\label{eq:defM}
\eeq
The product of deltas has to be antisymmetrized in the second indices as indicated by the notation. Explicitly, the first few of these operators are given by:
\beq
\begin{tabular}{r | l}
$n$ & $R_n$ \\
\hline
2 & $-  \tr  M^2$ \\
3 & $2  \tr  M^3 $ \\
4 & $-6  \tr   M^4 + 3  (\!\tr M^2)^2$ \\
5 & $24 \tr  M^5 - 20  (\!\tr M^2) (\!\tr  M^3)$ \\
\end{tabular}
\label{eq:Zn-examples}
\eeq
where e.g.~$\tr  M^3$ stands for the contraction $M_{\mu \nu} M_{\nu \lambda}M_{\lambda\mu}$ etc. Notice that $\tr  M=\Box\phi=0$.
Clearly, with this definition each of the operators $R_n$ is defined for any $d$, integer or non-integer.

However if $d<n$ the number of possible values of the $\nu$ indices on the RHS of \eqref{eq:Zn} is too small: for each choice of $\nu_1, \ldots, \nu_n$ there are necessarily duplicate indices and total antisymmetrization is not possible. This implies that each of the operators $R_n$ vanishes when $d=1,2,\ldots,n-1$.\footnote{This can also be shown by expressing the $R_n$ operators in terms of the eigenvalues of the matrix $M$ \cite{tcsa} or by applying the Cayley-Hamilton theorem \cite{Hogervorst:2015tka}. We thank Miguel Paulos for suggesting to us a proof using antisymmetrization.} These are, then, examples of evanescent operators---operators which are identically zero for certain integer dimensions, but nontrivial for non-integer dimensions. 

We hasten to add that $R_n$'s are not the only evanescents of the theory. For example, multiplying an evanescent $R_n$ by any other operator gives again an evanescent. More evanescents can be constructed applying the same basic antisymmetrization idea to other tensors. To give just one example: one can replace some of the $M$'s in \reef{eq:Zn} with $M'_{\mu\nu}=\del_\mu \phi\,\del_\nu\phi$. In subsection \ref{sec:gram} we will see a systematic approach to count and construct all the evanescent states below some cutoff in dimension, based on the Gram matrix. A natural conjecture, which we checked extensively, is that all the evanescents appear from antisymmetrizations of the above sort. It would be nice to prove this.

Let us finally note that $R_n$'s are not primary operators, as they can be represented as total derivatives:
\beq
R_n=\del_{\mu_1} \left\{ \delta^{\mu_1[\nu_1}\delta^{\vert\mu_2\vert\nu_2}
\dots\delta^{\vert\mu_n\vert\nu_n]} \no{ \del_{ \nu_1}\phi\, M_{\mu_2 \nu_2} \dotsm M_{\mu_n \nu_n}}\right\}\,.
\eeq
Below we will further analyze the conformal multiplets to which $R_n$ belong.

\subsection{Negative-norm states}
\label{sec:negnorm}

We will now establish that the Hilbert space of local operators of the
free scalar theory is not positive definite, and so the theory is not unitary, away from integer $d$.

Whether a Hilbert space is positive definite or not is encoded in the Gram matrix, or the matrix of inner products. In conformal field theory it's customary to work in radial quantization, where the inner products are evaluated inserting the operators at zero and infinity. The radial quantization Gram matrix is thus defined as
\be
G_{ij} \equiv \langle \OO_i | \OO_j \rangle_{\rm rad} \colonequals \lim_{x\to0}\langle [\calO_i(x)]^\dagger \calO_j(x) \rangle\,, 
\label{eq:Gdef}
\ee
where the conjugate operator $[\calO_i(x)]^\dagger$ is inserted at the point $Rx=x^\mu/x^2$ and is rescaled appropriately to have a finite limit. We will extensively use the above Gram matrix in the rest of the paper, and it will be discussed in more detail in Sec.~\ref{sec:gram}. 

For now, it will however be convenient to introduce a second inner product $\tilde{G}$. Here, one of the operators is inserted at the point N with coordinates $x_d = 1/2,$ $x_{1}, \ldots,x_{d-1} = 0$. Conjugate operators $[\Oo(N)]^+$ are inserted at the point S with coordinates $x_d = -1/2$, $x_1, \ldots, x_{d-1} = 0$, which is the reflection of N through the plane $x_d = 0$.\footnote{If $\Oo$ has non-zero spin, then $[\Oo(N)]^+$ also contains a factor $\Theta=(-1)^k$, where $k$ is the number of Lorentz indices perpendicular to the plane $x_d = 0$.} The Gram matrix $\tilde{G}_{ij}$ is then defined as
\be
\tilde G_{ij} \equiv \langle \OO_i | \OO_j \rangle_{\rm NS} \colonequals \langle [\calO_i(N)]^{+} \calO_j(N) \rangle\,.
\label{eq:Gtildedef}
\ee
This inner product is adapted to the so-called North-South (NS) quantization scheme~\cite{Luscher:1974ez,SRnotes}, where states in the Hilbert space are defined by inserting operators at N. 

The two Gram matrices $G$ and $\tilde{G}$ are related to each other, because there is a global conformal transformation which maps 0 and $\infty$ to N,S respectively. If there is a negative-norm state with respect to one norm, there will be one with respect to the other. Notice however that descendants transform non-homogeneously under this transformation, and so it's not true in general that $\tilde G_{ij}$ is proportional to $G_{ij}$ (it's only true for primaries).

Defining the norm via $\tilde G$ is more intuitive because of the relation to the reflection positivity in flat space, and also because the NS conjugation acts simply in the presence of derivatives. The $G$-norm is related to reflection positivity on the cylinder $S^{d-1}\times \mathbb{R}$, and its evaluation is slightly more complicated. In compensation it satisfies various nice selection rules which will be described in section \ref{sec:gram}. 

In this section we will exhibit operators in free massless scalar theory which have negative $\tilde G$-norm. The existence of such operators was first observed by us in \cite{tcsa}, and here we will give a slightly more detailed discussion. In the simplest and most explicit examples, the negative-norm operators will be Lorentz scalars. Recall that the two point function of a Hermitian scalar operator $\OO(x)$ (not necessarily a conformal primary) takes the form
\be
\langle \OO(x) \OO(0) \rangle = {c(\OO)}/{|x|^{2 \Delta_\OO}} \label{eq:2pt}\,,
\ee
where $\DD_\Oo$ is the scaling dimension of $\Oo$. 
For scalars the coefficient $c(\calO)$ is precisely the $\tilde G$-norm of $\calO$. This coefficient must be positive in a unitary CFT, while our operators will have it negative.\footnote{Since our understanding of field theory in non-integer dimensions is somewhat incomplete, one may wonder what precisely is meant by their `unitarity'. The conclusions of this paper are based on the natural assumption that $c(\OO) > 0$ is a necessary ingredient in any meaningful definition of a unitary theory.} 

If one wants to look for negative-norm states, the evanescent operators discussed in the previous section are good candidates. Indeed, their $\tilde{G}$-norms $c(\calO)$ will have zeroes at (some) integer $d$'s, and so assuming that these zeroes are first order, $c(\calO)$ will be negative on one side of each zero.

Consider thus the operators $R_n$. Since $R_n$ is identically zero for $d=1,2,\ldots,n-1$, we know that any correlation function $\expec{R_n(x)\ldots}$ will vanish for these $d$. In particular, this will be true for the two point function coefficient $c(R_n)$. Let us then write:
\be
c(R_n) =  \hat c(d,n) \prod_{i=1}^{n-1}(d-i)\,,
\label{eq:ckn}
\ee 
On general grounds, we can say that $\hat c(d,n)>0$ for all integer $d\ge n$, since the operator is then nontrivial, and the theory unitary. 
An explicit computation gives the following formula:\footnote{We have verified this formula  $n=2,3,4,5,6$ and conjecture that it holds for all $n$.}
\beq 
\hat{c}(d,n) = n! (n-1)  \left[d (n-1)+2 n \right] d^n (d-2)^n  (d-n+3)_n\,, 
\label{eq:hatc}
\eeq
where $(x)_n \equiv \Gamma(x+n)/\Gamma(x)$ is the Pochhammer symbol.\footnote{
\label{note:norm}We are working in the normalization in which $\expec{\phi(x)\phi(0)} = 1/|x|^{2\Delta_\phi}.$}

\begin{figure}[htbp]
\begin{center}
\includegraphics[scale=1.0]{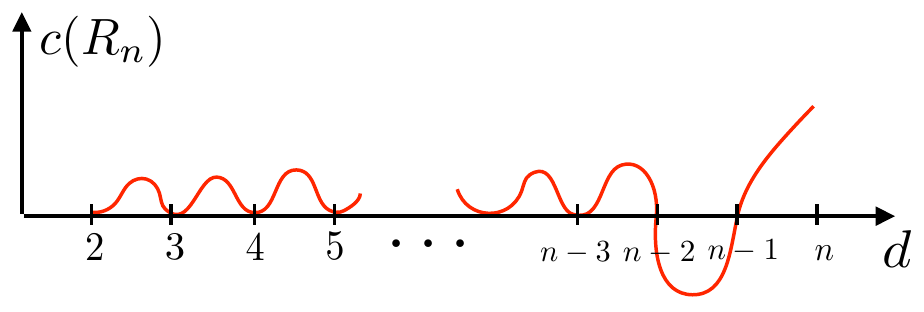}
\end{center}
\caption{The schematic behavior of $c(R_n)$, $n\ge 4$, as a function of $d$ in the range $d\ge2$. The norm vanishes for $d=2,\ldots,n-1$, is negative in the interval $n-2<d<n-1$, and is positive otherwise.}
\label{fig:cKn}
\end{figure}

We thus see that $\hat{c}(d,n)$ is positive at $d=n-1$, so that this zero of $c(R_n)$ remains first order. Consequently, $c(R_n)$ is negative for $n-2 < d < n-1$; see fig.~\ref{fig:cKn}. For example, $c(R_4)$ is negative for $2 < d < 3$, $c(R_5)$ is negative for $3 < d < 4$ etc. We conclude that the free massless scalar boson in any non-integer dimension $d>2$ contains negative-norm states and is therefore not unitary.\footnote{Notice the similarity with the proof that $O(N)$ models are not unitary for non-integer $N$ \cite{Maldacena:2011jn}.}

The somewhat complicated form of the $R_n$ operators explains why the negative-norm states have never been observed before our work \cite{tcsa}. We stress however that $R_n$ are nothing but linear combinations of normal-ordered products of the derivatives of $\phi$. They can be obtained by taking repeated OPEs of $M_{\mu\nu}$. In this sense these operators are not much different from, say, $\NO{\phi^2}$ or the stress tensor. In particular, there is no way to throw them out from the theory without destroying its consistency.

Above we discussed the two point function of the $R_n$ operators. Other correlators involving $R_n$ can be computed similarly. Just like the two point function, these correlators will vanish at integer $d=1\ldots n-1$ but will be generally nonzero for other $d$. Here is one example, for $n=5$, of a nonzero three point function:
\be
\langle \phi^2(x) \phi^3(y) R_5(0) \rangle = -240 (d-4) (d-3) (d-2)^6 d\, \frac{(d+2)x^2 y^2 - 3d (x\cdot y)^2}{|x|^{2+2d} |y|^{2+3d} }  \label{eq:3ptO}\,.
\ee
As expected, the three-point function vanishes at $d = 1,2,3,4$ when $R_5$ becomes null.\footnote{For $d = 1$ we have to use the identity $(x\cdot y)^2 = x^2 y^2$.} Notice that the three-point function does not have the standard Polyakov form of the correlator of three scalar primaries, because as we observed above $R_n$ is not a primary.

\subsection{Primary decomposition and negative norms}\label{sec:prim}

In the previous section we saw that the free massless scalar in $d>2$ contains negative-norm descendant states. What does this means in terms of primaries? Clearly, there should be either 

Scenario (1): negative-norm primaries above the unitarity bound, 

\noindent or 

Scenario (2): positive-norm primaries below the unitarity bound, 

\noindent or both. Recall that for primaries above the unitarity bound the norm of the descendants is related to that of the primary by a positive factor, so that if the norm of the primary is negative, so will be the norm of the descendant. For primaries below the unitarity bound there is at least one descendant whose norm has the sign opposite to that of the primary.

To decide between these scenarios, we would like to decompose the operators $R_n$ into a linear combination of descendants of primaries. 
In principle, it's a straightforward exercise to find such a decomposition for any given operator $\calO$. We start by acting on $\calO$ with $K_\mu$ several time until the result is zero. If, schematically, $(K_\mu)^N \calO=0$, then $(K_\mu)^{N-1} \calO$ is a linear combination of primaries, call them $\calO_i$ whose descendants $\calD_i$ contribute to $\calO$. Subtracting these descendants with appropriate coefficients, we get $(K_\mu)^{N-1} (\calO-c_i \calD_i)=0$. We can then move one level up and repeat the procedure recursively until the full decomposition is obtained. 

In practice however it's a bit tedious to carry this out for the operators $R_n$, which contain many derivatives, and one has to act many times with $K_\mu$ before hitting zero.
We performed the decomposition for $n\le 4$, and we will describe here the result for $R_4$. Its decomposition in general $d$ includes seven terms:
\begin{multline}
R_4= c_{1} \Box^4 \calO^{(1)}+c_2 \Box^2\del_\mu\del_\nu\calO^{(2)}_{\mu\nu} + c_3 \Box^2 \calO^{(3)}\\+ c_4\Box \del_\mu\del_\nu  \calO^{(4)}_{\mu\nu}
+c_5 \del_\mu\del_\nu\del_\lambda\del_\sigma \calO^{(5)}_{\mu\nu\lambda\sigma}+c_6\Box \calO^{(6)}+c_7 \del_\mu\del_\nu \calO^{(7)}_{\mu\nu}\,,
\end{multline}
where
\beq
\calO^{(1)} =\phi^4,\quad \calO^{(3)}= [(\del\phi)^2]^2+\ldots,\quad \quad \calO^{(6)}=(\del\phi)^2(\del\del\phi)^2+\ldots
\eeq
are scalar primaries (the $\ldots$ terms are fixed by the primary condition),
\beq
\calO^{(2)} =\phi^2 T_{\mu\nu}\, ,\quad \calO^{(4)}=\del_\mu\phi\,\del_\nu\phi(\del\phi)^2+\ldots,\quad \calO^{(7)}=\del_\mu\phi\,\del_\nu\phi(\del\del\phi)^2+\ldots
\eeq
are symmetric traceless spin two primaries (the first of these is a normal-ordered product of $\phi^2$ and the stress tensor), and
\beq
\calO^{(5)} =\del_\mu\phi\,
\del_\nu\phi\,
\del_\lambda\phi\,
\del_\sigma\phi+\ldots
\eeq
is a symmetric traceless primary of spin four. 

The $\tilde G$-norm of $R_4$ has a first order zero at $d=3$, and one may ask how this agrees with its decomposition. The above primaries are normalized so that they don't contain any explicit $d$-dependent factors which vanish or blow up at $d=3$.
In this normalization, the coefficients $c_1, \ldots, c_6$ turn out to be $O(d-3)$, hence the operators $\Oo^{(1)}  \ldots \Oo^{(6)}$ only give a subleading $O(d-3)^2$ contribution to the norm of $R_4$.
On the other hand $c_7$ remains finite as $d\to3$. Therefore, at leading order in $d-3$, the norm of $R_4$ is proportional to that of $\calO^{(7)}$. The proportionality coefficient is expected\footnote{
We say `expected' since strictly speaking the standard unitarity bounds have been derived for the integer dimensions. We rely on the plausible assumption that nothing unexpected happens with the unitarity bounds analytically continued to non-integer $d$. At any fixed level of the conformal multiplet, one could check this by an explicit computation, although we haven't done this.
}
to be positive, since $\calO^{(7)}$ is well above the spin two unitarity bound.
We therefore expect that the norm of $\calO^{(7)}$, as that of $R_4$, has a first order zero at $d=3$ and is negative at $d$ just below $3$. This corresponds to the above scenario (1). We computed the norm of $\calO^{(7)}$ explicitly and checked that this interpretation is indeed correct. We don't present the full details, except for the fact that $\calO^{(7)}$
can be written in the form
\beq
\calO^{(7)}_{\mu\nu}=const.\, \delta^{\mu_1[\nu_1}\delta^{\vert\mu_2\vert\nu_2}\delta^{\vert\mu_3\vert\nu_3} \delta^{\vert\mu_4\vert\nu_4]}(\phi M_{\mu_2,\nu_2}-9 \del_{\mu_2}\phi\, \del_{\nu_2}\phi) M_{\mu_3,\nu_3}M_{\mu_4,\nu_4}+O(d-3)\,,
\eeq
and we can use the same antisymmetrization argument as before to conclude that $\calO^{(7)}$ vanishes in $d=3$.

\subsection{The radial quantization Gram matrix}
\label{sec:gram}

In the previous sections we produced a few examples of negative-norm states in the free scalar theory. We would now like to give a more systematic survey, and in particular count how many states have negative norm. From now we will switch to using the radial quantization Gram matrix $G$ defined in \reef{eq:Gdef}. For the systematic analysis $G$ is more convenient than $\tilde G$ because it satisfies various nice selection rules (see below). Let's briefly review how $G$ is constructed in practice \cite{tcsa,Hogervorst:2015tka}. For us, the operators $\OO_{i}, \calO_j$ in \reef{eq:Gdef} will be normal-ordered products of $\phi$ and its derivatives, or linear combinations thereof. The action of conjugation on $\phi$ is defined by 
\beq
[\phi(x)]^\dagger := |x|^{-2\Delta_\phi} \phi(Rx)\,.
\eeq
It is extended to derivatives by (anti)linearity:
\beq
[\del_x^n \phi(x)]^\dagger := \del_x^n\bigl\{|x|^{-2\Delta_\phi} \phi(Rx)\bigr\}\,.
\eeq
It is also extended to the normal-ordered products of $\phi$ and its derivatives in an obvious way. With these definitions the Gram matrix will be Hermitian (in fact real and symmetric if the basis $\OO_i$ involves only linear combinations with real coefficients). Computing it is a problem of symbolic algebra, which we realized in {\sc Mathematica}. We will keep using the normalization from note \ref{note:norm}, in which $\langle \phi | \phi \rangle  = 1$. 

As mentioned, the radial quantization Gram matrix respects various selection rules, which give it a block structure. One basic rule is that only operators with the same $SO(d)$ symmetry and the same scaling dimensions have nonzero inner product. This is in fact true for any theory.

Our specific theory will have another, very powerful, selection rule. To describe it, define the \emph{$\phi$-type} of an operator as the list of the number of derivatives for each factor $\phi$ in the operator, sorted in increasing order. This list thus has length $n_\phi$ defined as the number of $\phi$'s in the operator. When determining the $\phi$-type, contractions are unimportant. For example, the $\phi$-type of $\phi^2$ is $\{0,0\}$ and the $\phi$-type of $\phi\, \del^2\phi\, \del \phi\, \del^3 \phi$, no matter how the indices are contracted, is $\{0,1,2,3\}$. Clearly, if two operators have the same $\phi$-type they also have the same scaling dimension, although the converse is not necessarily true.

With this definition, in the free massless scalar theory, a Gram matrix element vanishes unless the two operators are of the same $\phi$-type. A very simple partial case of this rule is that the matrix element vanishes unless $n_\phi(\calO)=n_\phi(\calO')$. This is clear since otherwise one cannot form any contraction between $\calO$ and $\calO'$, while the full rule may not appear totally obvious starting from the definition of $G$ in terms of two point functions.\footnote{See however the discussion in \cite{Hogervorst:2015tka}, section C.3.
}
This rule however directly follows from the oscillator picture of the free scalar Hilbert space in the canonical quantization, where each factor $\del^l\phi$ corresponds to adding an oscillator of angular momentum $l$; see \cite{tcsa}, Eq.~(3.13). 

The above discussion was for general $d$. Let us now describe how the Gram matrix can be used to enumerate the null states in 4d, which correspond to evanescent operators when we move slightly away from 4d. This will be also relevant for the subsequent discussion of the WF fixed point.

Using our {\sc Mathematica} code we evaluated all matrix elements for scalar parity-even operators up to $\Delta = 23$ in $d=4$.
This will be enough for the applications we have in mind. The code becomes slower for longer operators but with additional optimization one could go higher up in dimension if needed.

More precisely, we split the operators into groups having the same $\phi$-type. Within each group we computed the Gram matrix keeping the spacetime dimension $d$ as a free parameter. Setting $d$ to 4 and counting the number of zero eigenvalues we can then determine the number of null states. We then set $d$ to $4-\eps$ and determine the eigenvalues of the Gram matrix to first order in $\eps>0$. We found that up to the dimension that we are working all eigenvalues are either $O(1)$ or $O(\eps)$ as $\eps\to0$. In particular there are no $O(\eps^2)$ eigenvalues up to $\Delta=23$.\footnote{There is no contradiction between this statement and the fact that the $\tilde G$-norm of $R_7$ vanishes as $O(\eps^2)$ near $d=4$, see Fig.~\ref{fig:cKn}.
First of all, the $G$-norm discussed in this section is not proportional to the $\tilde G$-norm except for the primaries, and in fact the $G$-norm of $R_7$ is $O(\eps)$. Additionally, there is another evanescent operator of the same $\phi$-type as $R_7$, namely $\NO{R_2R_5}$. These two evanescents mix at $O(\eps)$, and the resulting eigenvalues are both $O(\eps)$.}
This fact will play some role later in the WF fixed point discussion. At still higher dimensions there might well be $O(\eps^2)$ eigenvalues.

Our results are summarized in Table~\ref{tab:nullstatescounting}. The first null state occurs at $\DD = 15$; this is the operator $R_5$. At $\Delta=16$ we have two null states: the operator $R'_5$ obtained by replacing one of the $M$'s by $M'$ in $R_5$, and the product $\NO{\phi R_5}$. In fact the normal-ordered product of any state with a null state is always a null state. The three null states at $\Delta=17$ are $\NO{\phi^2 R_5}$, $\NO{\phi R'_5}$ and $R''_5$ obtained by two $M\to M'$ replacements in $R_5$. A similar analysis can be done for $\DD > 17$. We see from the table that majority (although not all) of the null states acquire negative norm in $d=4-\eps$.

As an example, let's consider the block of scalar operators of $\phi$-type $\{2,2,2,2,2\}$. The space of operators of this $\phi$-type is two dimensional, with the basis naturally chosen as
\beq
\label{eq:OO12}
\Oo_1 = (\!\tr M^2)\, (\!\tr M^3) \quad \text{and} \quad \Oo_2 = \tr M^5.
\eeq
where we recall that $M_{\mu \nu} = \del_\mu \del_\nu \phi$. The Gram matrix restricted to this subsector is
\begin{multline}
\label{eq:exGram}
{\bf G}_{\{2,2,2,2,2\}} = 2 (d-2)^6 (d-1) d^2 (d+2) (d+4) \\
\times \begin{pmatrix} 12 d^2 \left(d^2+d+10\right) & 60 d \left(d^2+2 d-4\right) \\
 60 d \left(d^2+2 d-4\right) & 5 \left(d^4+7 d^3+12 d^2-48 d+96\right) \end{pmatrix}\,.
\end{multline}
Diagonalizing, we find two positive eigenvalues for $d>4$. One of them crosses zero and becomes negative for $d<4$. The corresponding eigenvector is in $d=4$ precisely the operator $R_5$.

\begin{table}[!htb]
\centering
\begin{tabular}{c|ccccccccc}
$\Delta$& 15 & 16 & 17 & 18 & 19 & 20 &
   21 & 22 & 23 \\
   \hline
$N_0$ & 1 & 2 & 3 & 6 & 12 & 24 & 46
   & 93 & 181 \\
$N_-$ & 1 & 2 & 3 & 5 & 10 & 21 & 41
   & 83 & 160 \\
\end{tabular}
\caption{\label{tab:nullstatescounting} $N_0$ is the number of scalar parity-even null states in $d = 4$ with scaling dimension $\Delta$. $N_-$ denotes how many of these null states become negative-norm slightly below four dimensions (the rest of the null states acquire positive norm). 
}
\end{table}

The number of physical states of dimension $\Delta$ in the massless scalar theory for integer $d$ can be counted using the oscillator representation \cite{Cardy:1991kr}; it grows as $\exp(C \Delta^{1-1/d})$.\footnote{The same growth rate is expected to hold in any CFT \cite{ElShowk:2011ag,Pappadopulo:2012jk}.} The number of null states will grow at least that fast, since we can form high-dimension null states by forming products of low-lying nulls with high-dimension normal states. It would be interesting to obtain a more systematic counting of the null states in any integer $d$. 

\subsection[General tensorial operators]{General tensorial operators\footnote{This section is somewhat outside the main line of development and may be skipped on the first reading, but it is helpful for the discussion in section~\ref{sec:synth}.}}\label{sec:tensops}

Our discussion so far was focused on scalar operators, with a brief mention of symmetric traceless tensor operators in subsection \ref{sec:prim}. This was clearly sufficient to demonstrate the non-unitarity of the free boson in non-integer dimensions, but for completeness let us now consider more general tensorial operators. 

As we mentioned above, the set of nontrivial local operators for the free boson in $d$ dimensions is generated by symmetrized normal ordered products of the form
\be
\label{eq:generaltensorop}
: \left( \partial_{\mu_1} \ldots \partial_{\mu_{n_1}} \phi\right) \left( \partial_{\nu_1} \ldots \partial_{\nu_{n_2}} \phi\right)  \ldots  \left( \partial_{\rho_1} \ldots \partial_{\rho_{n_k}} \phi\right) :
\ee
with $k \geq 1$ and $n_1 \ldots n_k \geq 0$, where some indices may be contracted modulo the sole constraint that $\square \phi = 0$. We would like to find a way to classify these operators into ``irreducible representations of the Euclidean rotation group in $d$ dimensions''. In order to do so we can (1) contract the indices with invariant tensors, and (2) symmetrize and/or antisymmetrize the uncontracted indices.

Let us first consider the invariant tensors. For any \emph{integer} $d$ there are precisely two fundamental invariants: the inverse metric $\delta^{\mu \nu}$ and the totally antisymmetric tensor $\epsilon^{\mu_{1} \ldots \mu_{d}}$. It is clear that our formalism for non-integer $d$ can include the former but not the latter. Indeed, the theory for non-integer $d$ should smoothly continue to \emph{any} integer $d$. By taking this integer $d$ as large as we want we can banish the epsilon tensor from appearing in any operator with finitely many $\phi$'s. We are therefore left with $\delta^{\mu \nu}$ as the sole fundamental invariant.

The prescription to classify the operators is then the following. We first remove all the traces from a general tensor, which can be done in arbitrary $d$ using $\delta^{\mu \nu} \delta_{\mu \nu} = d$. The indices on the resulting traceless operators are then symmetrized or antisymmetrized according to Young tableaux.\footnote{To obtain a nontrivial operator these tableaux must of course respect the permutation symmetries of the $\phi$'s as well as those of the derivatives on a single $\phi$. For example, the antisymmetric parts of both $\del_\mu \phi \del_\nu \phi$ and of $\del_\mu \del_\nu \phi$ clearly vanish.} This again can be done in a $d$ independent manner, with one important peculiarity: the absence of an epsilon tensor implies that there is no restriction on the number of rows in the tableau. In other words, all the different Young tableaux correspond to \emph{different} irreducible tensor structures in $d$ dimensions. It should be clear that antisymmetrization over $k$ indices gives an evanescent operator for any integer $d < k$, but a perfectly valid operator otherwise.

Finally we discuss the nontrivial reflection operator $P$ in the $O(d)$ group with determinant $-1$, which is the generalization of parity to arbitrary $d$. For our purposes we can define it to be a reflection in one of the $d$ coordinates, say coordinate number $1$. Under this transformation
\be
P \phi(x) P = \phi(P x) 
\ee
and therefore 
\be
P \del_{\mu} \phi(x) P  =  \del_{\mu} (\phi(P x)) = (-)^\mu (\del_\mu \phi)( P x)
\ee
where $(-)^\mu$ is shorthand for a sign that equals $-1$ if $\mu$ is the reflected index and $+1$ otherwise. This clearly extends to higher derivatives, and for any tensorial operator constructed in general $d$ as described above we have
\be
P T_{\mu_1 \ldots \mu_k} (x) P = (-1)^{\mu_1} \ldots (-1)^{\mu_k} T_{\mu_1 \ldots \mu_k} (P x)\,,
\ee 
irrespectively of how this tensorial operator is constructed from derivatives and $\phi$'s, and irrespectively of the symmetrization and trace properties of the indices. Operators with the above transformation properties are canonically defined as ``even'' under the generalized parity operation.

We conclude that in a formalism that can be defined for arbitrary $d$ there is no room for parity odd operators. Indeed, for the free scalar theory in integer dimensions a local operator is even/odd under parity precisely if we need an even/odd number of epsilon tensors to define the operator. Therefore, in our formalism the pseudoscalars are replaced with $d$-index antisymmetric tensors, the pseudovectors with $(d-1)$-index antisymmetric tensors, and so on: for every parity-odd operator in a given integer dimension there exists a parity-even operator that extends to arbitrary dimensions.\footnote{The precise recipe is the following: start with the definition of the operator using $\phi$'s, derivatives and invariant tensors. Now replace pairs of epsilon tensors with products of the metric, for example in two dimensions $\epsilon^{ab} \epsilon^{cd} = \delta^{ac} \delta^{bd} - \delta^{ad} \delta^{bc}$. Repeat until there is a single epsilon tensor. Simply cross that one out from the definition of the operator, to obtain a parity even operator with some newly freed indices. {\bf Note added:} A similar prescription is also sometimes used in multiloop QFT computations involving $\gamma^5$ matrices \cite{Larin:1993tq,Gracey:2000am}. We thank John Gracey for pointing this out to us.}

Let us finally remark that the selection rules coming from parity conservation in integer $d$ will now have to follow from ordinary Lorentz symmetry rules. As an example, the fact that the OPE of two scalars does not contain antisymmetric tensors also implies that no pseudoscalars appear.

\section{Review of the Wilson-Fisher theory at one-loop}
\label{sec:WF}

In the previous section, we have shown that the free massless scalar in non-integer $d$ is not unitarity. 
We will now turn our attention to the WF fixed point in $d=4-\eps$ dimensions with $\eps \ll 1$. 
It's pretty clear that this theory will also be not unitary. Indeed, its dynamics happens in the same non-positive-definite Hilbert space as that of the free theory. Barring unlikely coincidences and decouplings, negative-norm states found in the free theory will persist in the interacting theory. 
This fact and its consequences will be studied in detail in section~\phantom{,}~\ref{sec:nonunitWF}.

In this section, we will do some preparatory work, by reviewing the lowest-order computations of the WF critical exponents. This material is completely standard. Unitarity considerations will not play any role here. The main result is that the leading anomalous dimensions are computed by diagonalizing the matrix of the OPE coefficients of the operator $\phi^4$, evaluated in the free 4d theory, Eq.~\reef{eq:Gammafinal}. The reader familiar with this can proceed straight to section~\phantom{,}~\ref{sec:nonunitWF}.

\subsection{Anomalous dimension generalities}
\label{eq:WFsetup}
We will remind here some well-known facts about how one computes in perturbation theory scaling dimensions of local operators at the WF fixed point (or at any other weakly coupled IR fixed point). We will work in the minimal subtraction scheme. This is all standard, see e.g.~\cite{Kleinert,Vasiliev}.

One starts with bare operators $\calO_i$ which are the free theory operators from section \ref{sec:freetheory}. When one computes perturbative corrections to correlation functions of these operators, one encounters poles in $\eps$ which are subtracted order by order to define the finite operators denoted $[\Oo_i]$. The bare and finite operators are related by the mixing matrix $Z$ as follows: 
\beq
\label{eq:Zdef}
\Oo_i = Z\du{i}{j} [\Oo_j].
\eeq
Then $Z$ can be expanded order by order in $g$:
\beq
\label{eq:Zmat}
Z = \unit + g Z_1 + \ldots\,,
\eeq
where the omitted terms describe higher-order contributions in $g$. In minimal subtraction, the matrix elements $Z\du{i}{j}$ are fixed to be poles in $\eps$. 

Of course, the action~\reef{eq:WFlagrangian} also requires field and coupling constant renormalizations of order $g^2$ and higher. These effects ensure that the theory reaches a fixed point at $g=g_*$ which at leading order is given in \reef{eq:gcritical}. Apart from this, they will not play a role in our discussion.

The above renormalization prescription renders correlation functions of the operators $[\Oo_i]$ finite for any value of $g$. In particular, correlators of these operators at the IR fixed point can be computed by setting $g\to g_*$. 
Notice however that the operators $[\Oo_i]$ as defined above will usually not have well-defined IR scaling dimensions, i.e.~they will be mixtures of operators with different dimensions. To get operators with well-defined IR dimensions, an additional change of basis is needed. Namely, consider the dilatation operator acting in the space of operators:
\beq
D = D_0 + \Gamma(g_*)\,.
\label{eq:dilat}
\eeq
We split $D$ into the classical term $D_0$ and a matrix of anomalous dimensions $\Gamma$. 
We choose a basis of bare operators $\Oo_i$ that have a well-defined classical dimension $\DD_0$ in $d$ dimensions. In this basis $D_0$ is diagonal:\footnote{
This simply means that each of the basis operators have to be made of a fixed number of $\phi$'s and derivatives. E.g. $\phi^4$ and $(\del \phi)^2$ are good basis elements. However, since the dimensions of these operators are different at $O(\eps)$, a basis containing their linear combinations would not be a good basis for our purposes.
}
\beq
(D_0)\du{i}{j} = \DD_{0,i} \, \delta \du{i}{j}\,.
\eeq

By a standard RG argument, the anomalous dimension matrix $\Gamma(g)$ is related to the mixing matrix $Z(g)$ in \reef{eq:Zdef} as follows:
\beq
\frac{d}{d\log \mu} Z = Z \cdot \Gamma(g) \label{eq:Gammadef}
\eeq
where the derivative is taken holding the bare coupling $g_{\rm B} \equiv g \mu^\eps$ fixed. The renormalizability of the $\phi^4$ theory implies that $\Gamma$ depends only on $g$ and not on $\eps$.  

In terms of the dilatation matrix $D$, the operators which have well-defined scaling dimensions at the fixed point are linear combinations of finite operators of the form
\beq
c^i[\Oo_i]
\label{eq:goodop}
\eeq
where $c^i$ is a left eigenvector of the matrix $D$,
\be
c^i D\du{i}{j} = \Delta c^j\,.
\ee
The scaling dimensions themselves are the eigenvalues $\Delta$ corresponding to these eigenvectors. 

\subsection{$\Gamma$ at one-loop from the OPE}
\label{sec:WFopes}

To compute the scaling dimensions, we need the anomalous dimension matrix. At higher orders, the only currently known systematic way of finding it is the full-blown Feynman-diagrammatic perturbation theory. In this paper we will only need $\Gamma$ at one loop. There is then a well-known simple alternative (see e.g.~\cite{Cardy}) which instead of Feynman diagrams uses the position space OPE, and which we will now review. For us this method is more convenient, since we have a well-developed machinery to compute the OPEs efficiently, described in detail in Ref.~\cite{tcsa}.

To avoid possible misunderstanding, we would like to stress the difference between the approach adopted here and the recent work~\cite{Rychkov:2015naa,Basu:2015gpa,Raju:2015fza,Ghosh:2015opa,Sen:2015doa} on how to compute the dimensions of operators at the WF fixed point in the $\eps$-expansion without any reference to the Lagrangian but just from the constraints of conformal symmetry. That work is interesting and radical although still in its infancy. Here we will be not at all radical: we will be relying as usual on the UV Lagrangian and the path integral to define the theory and do the computation, just phrase it in the language of the OPE rather than Feynman diagrams. 

Let us then compute the one-loop matrix $Z$ using the OPE method. We will determine the matrix $Z_1$ in \reef{eq:Zmat}; the matrix $\Gamma$ will then follow. As is well known, in minimal subtraction operators can mix only if they have the same 4d dimension. The matrix $Z$ is therefore block diagonal, with each block corresponding to operators with the same 4d dimension. We call these ``big blocks'' to distinguish them from smaller blocks that will be introduced below. Within the big block, the classical dimensions in $d=4-\eps$ will differ by $O(\eps)$, and it will be convenient to introduce an antisymmetric matrix $\kappa_{ij}$ measuring this difference:
\beq
\Delta_{0,i}-\Delta_{0,j}=\kappa_{ij}\eps\,.
\eeq
Within a big block, we write the one-loop matrix $Z_1$ as
\beq
\label{eq:Z1def}
(Z_1)\du{i}{j} = \frac{1}{\eps}\mu^{\kappa_{ij}\eps} z\du{i}{j}\,,
\eeq
where the power of $\mu$ is dictated by dimensional analysis, and $z$ is an order one numerical matrix which we have to compute.

Suppose that we want to determine the one-loop counterterm for a generic scalar operator $\calO_i$. Consider the free theory OPE of $\phi^4$ with this operator:\footnote{We do not distinguish between conformal primaries and descendants here.} 
\beq
\phi^4(x) \calO_i(0) = \sum_j {C\du{i}{j}}{|x|^{-(\Delta_0(\phi^4)+\Delta_{0,i}-\Delta_{0,j})}} \calO_j(0) \; + \; \text{non-scalar operators}. \label{eq:OPE}
\eeq 
Here $\Delta_0(\phi^4)=d-\eps$ is the classical dimension of the operator $\phi^4$. The $C\du{i}{j}$ is the free theory OPE coefficient matrix. For simple operators it's easy to compute by counting Wick contractions. For more complicated operators it makes sense to use symbolic algebra, as we will do here following \cite{tcsa}. 

Now, the matrices $C$ and $z$ can be related by the following path integral argument. The leading correction to any free theory correlation function $\expec{\Oo_i(0)\ldots}_\text{free}$ is given by
\beq
-\frac{g\mu^{\eps}}{4!}\int\! \mathrm{d}^dx\, \expec{\Oo_i(0) \phi^4(x)\ldots}_\text{free}\,.
\label{eq:pertg}
\eeq
The pole in $\eps$ that we need to subtract will come from the short-distance part of the integration region, $x \to 0$. In this region the OPE is valid, and we can replace $\Oo_i(0)\phi^4(x)$ by the RHS of Eq.~\reef{eq:OPE}. Since the $x$-integral is spherically symmetric, it picks out only  the scalars, and gives rise to a factor of $\Sd = 2\pi^2 + O(\eps)$.  Next, we integrate over the radial direction. If the result is order $1/\eps$, then it will have to be compensated by counterterms. This occurs only if the operator $\Oo_j$ is within the same ``big block'' as $\calO_i$---a manifestation of the mentioned fact that only operators with the same 4d dimensions mix. Doing the radial integral, the pole part is then given by
\beq
\label{eq:sing}
-\frac{g}{4!}\frac{2\pi^2}{(1-\kappa_{ij})\eps} C\du{i}{j} \expec{\Oo_j(0)\ldots}_\text{free}\,,
\eeq
where in the rhs we have a sum over all $j$ in the big block of $i$.

To be a bit more precise, the OPE coefficients $C\du{i}{j}$ depend on $\eps$, and the actual residue of the $1/\eps$ pole in \reef{eq:sing} is obtained by replacing $C\du{i}{j}$ with
\beq
\overline{C}\du{i}{j} \equiv C\du{i}{j}\big|_{d = 4}\,. \label{eq:Cbardef}
\eeq
We cancel the pole by adding to $\calO_i$ the corresponding counterterm, defining
\beq
[\calO_i]=\calO_i+g\mu^{\kappa_{ij}\eps}\frac{\pi^2/12}{(1-\kappa_{ij})\eps} \overline{C}\du{i}{j}\calO_j\,.\label{eq:counterterm}
\eeq
Notice that the power of $\mu$ in the RHS of this expression is not given by the above computation but is reconstructed by dimensional analysis. Comparing to Eq.~\reef{eq:Zdef}, we obtain our final result:
\beq
z\du{i}{j} = \begin{cases}\displaystyle-\frac{\pi^2/12}{1-\kappa_{ij}} \overline{C}\du{i}{j}\,,&i,j\in\text{same big block,}\\
0\,,& \text{otherwise.}\,
\end{cases}
\label{eq:zC}
\eeq
To summarize: to compute the one-loop mixing matrix $Z_1$, it's enough to evaluate the OPE with the operator $\phi^4$. Moreover, we only need a small part of this OPE, namely the part which involves scalars with the same 4d dimension.

Notice that in deriving \eqref{eq:sing} we needed to assume that $\kappa_{ij} \ne 1$, because otherwise the integral is logarithmic and is not rendered finite in dimensional regularization. If this occurred it would be paradoxical, signaling a breakdown of the minimal subtraction scheme. However, this does not happen, because it can be shown that $\overline{C}\du{i}{j}=0$ for all such pairs of operators.

In fact, an even stronger statement holds (see appendix~\ref{sec:selecRule} for the proof): for operators with the same 4d dimension, $\overline{C}\du{i}{j}$ is zero unless $\kappa_{ij}= 0$. An equivalent form of the latter condition is that $\calO_j$ and $\calO_i$ contain the same number of $\phi$'s.
The set of all operators which have the same 4d dimension and the same number of $\phi$'s will be called the ``small block" of $\calO_i$. In other words, we are saying that at one loop operators mix only within the small blocks. This fact was already noticed in Ref.~\cite{Kehrein:1992fn}.  Examples show that this rule does not extend to higher orders in perturbation theory.\footnote{For example $\phi^4$ mixes with $\del^2(\phi^2)$ at $O(g^2)$.}

Using \reef{eq:Gammadef}, \reef{eq:Z1def}, \reef{eq:zC}, the one-loop anomalous dimension matrix within a given big block is given by
\beq
\Gamma\du{i}{j}(g) = g(-1+ \kappa_{ij}) z\du{i}{j}=g \frac{\pi^2}{12} \overline{C}\du{i}{j}\,\quad(\text{canonical normalization}).
\label{eq:Gamma}
\eeq
The specification in parenthesis reflects that in the above discussion we used the OPE coefficients computed using the canonically
normalized scalar field as in \reef{eq:WFlagrangian}. We will now pass to the CFT normalization of note \ref{note:norm} used elsewhere in the paper.
The two normalizations are related by:
\beq
\phi_{\text{can}}=k_d^{1/2}\,\phi_{\text{CFT}},\qquad k_d=1/[(d-2)\Sd].
\eeq
For the matrix of OPE coefficients of the $\phi^4$ operator this introduces an extra factor $k_d^2$, which in 4d becomes $k_4^2 = 16 \pi^4$. Taking into account this factor and substituting the critical coupling \eqref{eq:gcritical}, we find:
\beq
\label{eq:Gammafinal}
\Gamma\du{i}{j}(g_*)=\frac{\eps}{36} \,\overline{C}\du{i}{j}\, \quad(\text{CFT normalization}),
\eeq
which is our final master formula for the one-loop anomalous dimension matrix. To avoid any possible confusion, we emphasize that this formula is valid when operators $i,j$ are within the same big block. In addition, within a given big block $\overline{C}\du{i}{j}=0$ unless the two operators are in the same small block. For $i,j$ not in the same big block we have $\Gamma\du{i}{j}=0$ no matter what $\overline{C}\du{i}{j}$ is (it may be nonzero in this case).

\subsection{Examples}
\label{sec:ex}

Here are a few examples of anomalous dimensions computed in this formalism. Let's consider scalar operators that are even under the $\mbb{Z}_2$ symmetry $\phi \to - \phi$. The OPEs will be written schematically, indicating the coefficient $\overline{C}_i{}^j$ in the CFT normalization but leaving the dependence on $x$ implicit. Nonscalar operators and operators from different big blocks will be omitted. With $\gamma_\Oo$ we denote the eigenvalue of the anomalous dimension matrix $\Gamma(g_*)$ corresponding to the operator $[\Oo]$.

For the $\phi^2$ operator we have:
\beq
\phi^4\times \phi^2 = 12\,\phi^2\,.
\eeq
We recover the known result for the leading anomalous dimension $\gamma_{\phi^2}=\eps/3$.

We next consider the big block of operators of 4d dimension $4$, whose basis is formed by $\phi^4$ and $\del^2 \phi^2$. The OPE turns out to be diagonal:
\beq
\phi^4\times \phi^4 = 72\,\phi^4,\qquad \phi^4\times \del^2 \phi^2 = 12\,\del^2 \phi^2\,.
\eeq
This is in agreement with the small block selection rule.
The leading anomalous dimension $\gamma_{\phi^4}=2\eps$ agrees with the well-known result, while for $\del^2 \phi^2$ we find $\gamma=\gamma_{\phi^2}$ as it should be.

The big block of dimension $6$ consists of $\phi^6$, $\del^2 \phi^4$ and $\del^2\del^2\phi^2$. The $\phi^4$ OPE coefficients are again diagonal: 180, 72, 12, and there is only one new anomalous dimension.

The situation at dimension $8$ is more interesting. The big block has six operators. Three operators, $\phi^8$, $\del^2 \phi^6$ and $\del^2 \del^2\del^2 \phi^2$, form three small blocks by themselves. The $\phi^4$ OPE matrix in this sector is diagonal: 336, 180, 12. We get one new eigenvalue as expected. The three remaining operators have 4 derivatives and 4 $\phi$'s each, and thus form a separate small block. In the basis
\beq
\calA_1=\phi^2(\del_\mu\del_\nu \phi)^2,\quad 
 \calA_2=\phi(\del_\mu\del_\nu\phi)(\del_ \mu\phi)(\del_\nu \phi),
 \quad \calA_3=(\del \phi)^2 (\del\phi)^2\,,
\eeq
the $\phi^4$ OPE matrix in this sector is:
\beq
\left(
\begin{array}{ccc}
 56 & 32 & 0 \\
 4 & 60 & 6 \\
 0 & 16 & 48 \\
\end{array}
\right)\,.
\eeq
The eigenvalues are 72, 52, 40, and the corresponding left eigenvectors are:
\beq
c_1=(1,4,1),\quad
c_2=(-2,2,3),\quad
c_3=(1,-4,3)\,.
\eeq
In fact, the operator $c_{1}\cdot \calA$ is proportional to $\del^2\del^2\phi^4$ and indeed has the same anomalous dimension.
The other two linear combinations, $c_2 \cdot\calA$ and $c_3\cdot \calA$, are not derivatives of previously considered operators, and give two new anomalous dimensions. The two eigenvalues and eigenstates we report are in agreement with Ref.~\cite{Kehrein:1994ff}.

\renewcommand\thesection{\!\!\!\!\arabic{section}$-\epsilon$} 
\section[Non-unitarity at the Wilson-Fisher fixed point]{Non-unitarity at the Wilson-Fisher fixed point}
\renewcommand\thesection{\arabic{section}}
\label{sec:nonunitWF}

In the previous section we reviewed the standard facts about the leading anomalous dimensions at the WF fixed point. The considered examples concerned low-dimension operators. We know from section \ref{sec:freetheory} that once we move to sufficiently high dimension we will encounter evanescent operators and the related issue of non-unitarity. This raises several questions, such as: 
\begin{itemize}
\item Do the negative-norm states persist when we pass from the free theory to the WF fixed point? 
\item Should the above recipe for computing anomalous dimensions be modified in presence of evanescent operators/negative norms?
\item Is there any smoking-gun consequence of the non-unitarity at the level of the anomalous dimensions?
\end{itemize}
These and related questions will be discussed in this section.

\subsection{Robustness of negative-norm states}

One potentially confusing issue is as follows. If we write evanescent operators in the most natural parametrization in terms of $\phi$ and its derivatives, without any $d$-dependent factors, like in the examples considered in section \ref{sec:freetheory},
their norm vanishes for $\eps\to0$. This may create a feeling that the evanescent operators are ``fragile'', in the sense that even small perturbative corrections could dramatically change their properties, e.g.~change the sign of their norm, or perhaps decouple them completely from the theory. For the same reason one could worry that one has to be especially careful when computing their anomalous dimensions, and that the recipe from setion \ref{sec:WF} would require modifications. As we will now explain, this is not the case.

Basically what happens is that although correlators involving evanescent operators are small (order $\eps$), corrections to them are still one power of $\eps$ smaller. Indeed, these corrections involve correlators of the same evanescent operator with insertions of $\phi^4$, which are again order $\eps$, times the coupling $g$ which provides an extra suppression; see \reef{eq:pertg}.

A more systematic way to explain the same is to observe that the magnitude of the norm of an operator is unphysical, and once it is nonzero we can always rescale the operators to have order one positive or negative norm. This would require multiplying evanescent operators by a factor $1/\sqrt{\eps}$. 
After the rescaling all Gram matrix eigenvalues will be $O(1)$. In detail, the Gram matrix before the rescaling has the schematic form:
\beq
G_{ij}\sim \begin{pmatrix} 1 &\eps \\
\eps &  \eps \end{pmatrix}\,,
\label{eq:Gorig}
\eeq
where we indicate the order of magnitude of the matrix elements; the first column stays for the normal states, whose norm remains $O(1)$ as $\eps\to0$, and the second for the evanescents. After the rescaling this becomes:
\beq
G^R_{ij}\sim \begin{pmatrix} 1 & \sqrt{\eps} \\
\sqrt{\eps} & 1 \end{pmatrix}\,.
\label{eq:Gresc}
\eeq
We see that at leading order in $\eps$, the mixing between the two groups of states, described by the off-diagonal matrix elements, is a subleading effect. The orthonormal basis of the Hilbert space can be determined ignoring this mixing and diagonalizing the Gram matrix separately in the normal and evanescent sectors. It's important for this last statement that all eigenvalues in the evanescent sector be $O(\eps)$ and not higher order. As mentioned in section \ref{sec:gram}, this is true up to the dimension that we are working. 

Let us now apply the same rescaling to the OPE matrix $C\du{i}{j}$. Before rescaling we have
\beq
C\du{i}{j}\sim \begin{pmatrix} 1 &1  \\
\eps &  1 \end{pmatrix}\,,
\label{eq:Cbefresc}
\eeq
Here we took into account that if $\calO_i$ is an evanescent, all operators appearing on the rhs of \reef{eq:OPE} are necessarily evanescents. This is because the three point functions 
$\langle \text{(evanescent)}\phi^4 \text{(any)}\rangle$ have to vanish as $d\to4$. Thus we have a selection rule $C\du{\text{evanescent}}{\text{normal}} =O(\eps)$, reflected in \reef{eq:Cbefresc}. After rescaling, paying attention to the lower and upper indices, we have:
\beq
C^R_{i}{}^{j}\sim \begin{pmatrix} 1 &\sqrt{\eps} \\
\sqrt{\eps} &  1 \end{pmatrix}\,.
\label{eq:Caftresc}
\eeq
This has the same structure as the rescaled Gram matrix \reef{eq:Gresc}. We conclude that at leading order the anomalous dimensions in the evanescent and the normal operator sectors can be computed independently. Mixing between the two sectors is a higher order effect. Leading anomalous dimensions in the evanescent sector are generically $O(\eps)$, just as in the normal sector. They are described by the order-$\eps$ master formula \reef{eq:Gammafinal}, where $\overline{C}\du{i}{j}$ is the OPE matrix in the evanescent sector with $\eps$ set to zero.

When we work in the rescaled frame, the distinction between the normal and evanescent operators fades away (except that some evanescent operators have negative norm). For the normal operators, the most important effect of turning on the interaction is to induce anomalous dimensions which change the long-distance scaling of the correlation functions. On the other hand the numerical coefficient of the two point correlation function obtains only finite corrections of relative size $O(\eps)$ between its UV and IR values. This is manifest in the standard solution of the Callan-Symanzik equation in terms of the running coupling. This argument does not depend on the sign of the coefficient and applies also to the negative-norm states. We conclude that the states which had negative norm in the UV will remain negative-norm at the WF fixed point for $\eps\ll 1$. 
The negative-norm states are robust.
 
\subsection{Complex anomalous dimensions}
\label{sec:complex}
Unitarity puts well-known constraints on the spectrum of scaling dimensions in CFTs \cite{Mack:1975je,Minwalla:1997ka}. In particular, the spectrum of scaling dimensions in a unitary CFT is always real and bounded from below. One expects that nonunitary interacting theories should generically violate these properties, although sometimes they are preserved even without unitarity.\footnote{One example is the $i\phi^3$ Lee-Yang critical point in $2\le d<6$ which is $PT$-symmetric and the spectrum is real, although the unitarity bounds are violated. For $d=2$ this is the $\calM_{2,5}$ minimal model.}

In this section we will show that, as a dramatic repercussion of non-unitarity, the WF fixed point contains states whose dimensions become complex at order-$\eps$. 

First of all let's see how the complex eigenvalues can appear in our language.
As explained above, the order-$\eps$ dimensions are computed by solving the eigenvalue problem
\beq
v^i \, \overline{C}\du{i}{j} = \la \, v^j\,.
\label{eq:ev}
\eeq
This can be solved in each small block separately, as different small blocks don't mix. 
We can also multiply both sides by the Gram matrix and obtain the generalized eigenvalue problem
\beq v^i \, \overline{C}_{ij} = \la\, v^i\,  G_{ij}\,. \eeq
Here the matrix appearing in the lhs is the matrix of three point functions:
\beq \overline{C}_{ij} \equiv \overline{C}_{i}{}^{k} G_{kj} = \frac{1}{\Sd} \int_{|x|=1}\!\mrm{d}^dx\, \braket{i}{\phi^4(x)}{j}\,.\eeq
As such it is a real and symmetric matrix in our basis, while $\overline{C}_{i}{}^{j}$ did not have any symmetry properties.
Notice that the Gram matrix $\phi$-type selection rule is stronger than the small block selection rule, and so the small blocks remain decoupled even in the generalized eigenvalue problem formulation. 

Now, if the theory is unitary then the Gram matrix is positive definite. It can then be orthonormalized, reducing the generalized eigenvalue problem to a standard one which can have only real eigenvalues. Thus the anomalous dimensions will be real in unitary theories. 

In our problem, the normal and evanescent sector are decoupled at order $\eps$, and the normal sector is unitary. Thus the anomalous dimensions in the normal sector are all real.

Turning to the evanescent sector, it contains negative-norm states, so there is potential for complex eigenvalues. Notice however that if a decoupled sector contains \emph{only} negative-norm states, the eigenvalues will be again real by the same argument as above.

On the other hand generalized eigenvalue problems with both negative- and positive-norm states can lead to complex eigenvalues, provided the off-diagonal mixing is sufficiently large. As the simplest 2$\times$2 example, consider
\beq
\overline{C}_{ij} = \begin{pmatrix} 1 & \sigma \\ \sigma & 1 \end{pmatrix}\,,\quad G = \text{diag}\{1,-1\}\,,
\eeq
which has two eigenvalues $\pm (1-\sigma^2)^{1/2}$, complex for $\sigma>1$. Complex eigenvalues, if they exist, will always form complex-conjugate pairs as in this example.

\begin{table}[!htb]
\centering
\begin{tabular}{|c|c|c|cc|cc|ccc|ccc|}
\hline
$\Delta$ & 15 & 16 & 17 & 17 & 18 & 18 &
   19 & 19 & 19 & 20 & 20 & 20
   \\
   \hline
$n_\phi$ & 5 & 6 & 5 & 7 & 6 & 8 & 5 & 7
   & 9 & 6 & 8 & 10 \\
$N_0$  & 1 & 2 & 1 & 2 & 4 & 2 & 3 & 7
   & 2 & 14 & 8 & 2 \\
$N_+$ & 0 & 0 & 0 & 0 & 1 & 0 & 0 & 2
   & 0 & 1 & 2 & 0 \\
$N_-$ & 1 & 2 & 1 & 2 & 3 & 2 & 3 & 5
   & 2 & 13 & 6 & 2 \\
   \hline
   \end{tabular}
\\[5pt]
   \begin{tabular}{|c|cccc|cccc|ccccc|}
   \hline
 $\Delta$ & 21 & 21 & 21 & 21 & 22 & 22 &
   22 & 22 & 23 & \bf 23 & 23 & 23
   & 23 \\\hline
$n_\phi$ & 5 & 7 & 9 & 11 & 6 & 8 & 10 &
   12 & 5 & \bf 7 & 9 & 11 & 13 \\
   $N_0$ & 7 & 29 & 8 & 2 & 43 & 40 & 8
   & 2 & 17 & \bf 110 & 44 & 8 & 2\\
$N_+$  & 0 & 3 & 2 & 0 & 3 & 5 & 2 & 0
   & 0 & \bf 13 & 6 & 2 & 0 \\
$N_-$ & 7 & 26 & 6 & 2 & 40 & 35 & 6
   & 2 & 17 & \bf 97 & 38 & 6 & 2\\
   \hline  \end{tabular}
\caption{\label{tab:detailed}Distribution of the evanescent states from table \ref{tab:nullstatescounting} into small blocks characterized by a given $n_\phi$. Complex eigenvalues first occur in the small block shown in boldface.}
\end{table}

In table \ref{tab:detailed}, which is a more detailed version of table \ref{tab:nullstatescounting}, we show the distribution of evanescent states in small blocks. Recall that for a given $\Delta$ the small blocks are characterized by $n_\phi$---the number of $\phi$'s inside the operator. We see that the first small block with both negative- and positive-norm evanescents is the one at $\Delta=18$ with $n_\phi=6$. However the off-diagonal mixing turns out to be insufficient, so the anomalous dimensions are still real. This situation persists until $\Delta=22$.\footnote{Although this is not essential, there is another circumstance which contributes to the rarity of complex anomalous dimension. We will demonstrate it by an example. Take the small block with $\Delta=20$, $n_\phi=6$, which contains 1 positive- and 13 negative-norm evanescents, so one could hope to find complex eigenvalues here. However in fact four of these states will be of the form $\del^2$ acting on the evanescents in the $\Delta=18$, $n_\phi=6$ small block, and thus have the same anomalous dimensions, which as we said are real because the off-diagonal mixing is small. Thus we only have 10 really new evanescents in the $\Delta=20$, $n_\phi=6$ small block, and they are all negative-norm, so their anomalous dimensions are bound to be real.} Finally, at $\Delta=23$ we find complex anomalous dimensions, which appear in the small block with $n_\phi=7$. 

This finding is important and we will document it here in detail. Here is how the computation proceeds:

1. Compute the Gram matrix of dimension $\DD = 23$ parity-even scalar operators in $d=4$, and identify all the null states. The null states are zero-eigenvalue eigenvectors $v_{(a)}^{i}$ of the 4d Gram matrix
\beq
\overline{G}_{ij}=G_{ij}|_{d=4}
\eeq
Here $i$ runs over all $\DD = 23$ basis states, and $a$ numbers the null states. 
This step is simplified by the fact that the Gram matrix satisfies the $\phi$-type selection rule, and is block-diagonal. There are 2814 operators at $\Delta=23$, and the largest Gram matrix block has size $97$, corresponding to the $\phi$-type \{1,1,2,2,3,3,4\}.\footnote{This 97 is not the same as the boldface 97 in table \ref{tab:detailed}---in fact this block contains only 13 null states.}
 
2. The previous step results in 181 null states for $\DD = 23$, which split into five small blocks (see table \ref{tab:detailed}). The $n_\phi=7$ small block contains 110 states. Although this is not strictly speaking necessary, let us check that it has both positive and negative-norm states when we move away from 4d, which is the necessary condition to find complex dimensions.

The $(4-\eps)$-dimensional Gram matrix restricted to the evanescent sector is
\beq
\mathcal{G}_{ab}=v_{(a)}^{i}G_{ij}v_{(b)}^{j}\,.
\eeq
Here in the rhs we have the $(4-\eps)$-dimensional $G$ but the 4d null state eigenvectors. I.e.~we just use the same 4d null vectors $v_{(a)}^{i}$ as a basis of evanescent states in $(4-\eps)$-dimensions.\footnote{There is arbitrariness in how to extend the null state from 4d, since we can always add any other state with an explicit $\eps$-factor, and it will still be an evanescent. It's easy to see that this arbitrariness does not affect the results.} The so defined matrix $\mathcal{G}_{ab}$ is the lower-right corner in the schematic Gram matrix \reef{eq:Gorig}. We can then diagonalize $\mathcal{G}_{ab}$ expanded to first order in $\eps$, and count how many of its eigenstates have positive and negative norm. In the $n_\phi=7$ small block we find 13 positive- and 97 negative-norm states below 4d, as reported in the table. 

3. To find the anomalous dimensions, we have to diagonalize the matrix $\overline{C}_{i}{}^{j}$ within each small block. As explained above, the normal and evanescent operators decouple at order-$\eps$. Anomalous dimensions of the evanescent operators are computed by diagonalizing the matrix $\overline{C}_{i}{}^{j}$ restricted to the corresponding subspace. This is the lower-right corner matrix in the schematic equation \reef{eq:Cbefresc}. This matrix can be isolated in several ways; one that we used is as follows. Take a 4d null operator described by the vector $v_{(a)}^{i}$ and act on it with the 4d OPE matrix $\overline{C}\du{i}{j}$ (which we know). We know that we should get a linear combination of evanescents. Thus we simply solve the linear equation
  \beq
  v_{(a)}^{i} \, \overline{C}\du{i}{j} = \mathcal{C}\du{a}{b} \, v_{(b)}^{j}\,.
  \eeq
  for an unknown matrix $\mathcal{C}\du{a}{b}$.

The order-$\eps$ anomalous dimensions of the evanescents are the eigenvalues of $\mathcal{C}\du{a}{b}$. For the $\Delta=23$, $n_\phi=7$ sector it's a $110\times110$ matrix with rational coefficients which we do not report here. The matrix can be diagonalized numerically and one sees two complex-conjugate eigenvalue pairs:
\begin{align}
\la \approx  16.93372103 \pm 5.59469106i,\ 42.88540243 \pm 1.07557547 i\,.
\end{align}
 As a matter of fact the characteristic polynomial of $\mathcal{C}\du{a}{b}$ has a degree 26 factor which is irreducible over $\mathbb{Q}$ and has above complex eigenvalues as its roots. The expression for this factor is not particularly illuminating; it is included as a comment in the TeX source file of this paper.

We conclude that there are two pairs of scalar operators in the Wilson-Fisher fixed point in $4-\eps$ dimensions with complex-conjugate IR scaling dimensions. By the master formula \reef{eq:Gammafinal}, their dimensions are related to the above $\lambda$'s by
\beq
\Delta = 23 -\frac{7}{2}\eps + \gamma\,,\qquad\gamma = \frac{\la}{36}\eps + O(\eps^2)\,.
\eeq
We have checked that the eigenoperators corresponding to these complex dimensions can be represented as total derivatives and thus are not primaries. It should be possible to determine the primaries from which they originate acting along the lines of Sec.~\ref{sec:prim}, but we have not done this exercise.

\subsection{Comparison to the work of Kehrein et al.}
\label{sec:Kehrein}
The spectrum of Wilson-Fisher fixed points in $4-\eps$ dimensions has been previously studied in a series of remarkable papers by Kehrein et al.~\cite{Kehrein:1992fn,Kehrein:1994ff,Kehrein:1995ia} Here we would like to review their work and to compare with our results. In particular we would like to explain why they have not observed any complex dimensions nor negative-norm states.

They considered the general $O(N)$ case, but we will specialize to $N=1$ in this comparison. We will focus on the first two papers~\cite{Kehrein:1992fn,Kehrein:1994ff} which were devoted to the one-loop anomalous dimensions.\footnote{The third paper \cite{Kehrein:1995ia} studied two-loop anomalous dimensions for $N>1$ and for a particular subclass of operators, transforming as symmetric traceless tensors in all internal and derivative indices (i.e. without any $O(N)$ or Lorentz index contractions).}

Kehrein et al.\@ build their composite operators multiplying the elementary ``letters"
\beq
\Phi^{(l,m_1,m_2)}=h^{(m_1,m_2)}_{\alpha_1\ldots\alpha_l}\del_{\alpha_1}\ldots\del_{\alpha_l}\phi\,\,,
\eeq
obtained by contracting the derivatives of $\phi$ with constant symmetric traceless rank-$l$ tensors. These tensors $h^{(m_1,m_2)}_{\alpha_1\ldots\alpha_l}$, $m_1,m_2=-\half l\ldots \half l$ were chosen to form a basis for the $(\half l,\half l)$ representation of $so(4)=so(3)\oplus so(3)$.
This is in 4d, and when they move to $d=4-\eps$, they continue using the same 4d operator basis. In fact, $\eps$ enters in their computation only to produce the $1/\eps$ pole in the one-loop integral, and to supply the value of the fixed point coupling. 

Since Kehrein et al.\@ work only with 4d tensors, it is not surprising that they have not observed the evanescent operators whose very existence is due to peculiarities of tensor algebra in non-integer dimensions. A fortiori they could not observe non-unitarity and complex dimensions. Their way of proceeding amounts to restricting from the start to the subspace of normal operators, and neglecting the evanescents. Within this subspace, their results about order-$\eps$ anomalous dimensions are correct and in agreement with ours (see e.g.~examples in section \ref{sec:ex}).\footnote{They compute their anomalous dimension matrix in a rather different-looking, but in fact equivalent, way.}

\subsection{Higher order effects}

Given that at order-$\eps$ the evanescent operators decouple from the normal ones, it may be tempting to dismiss them as a curiosity which will never affect anything physically relevant. This would be wrong. In fact, already at order $\eps^2$ the effects of evanescent operators are expected to change the anomalous dimensions of physical operators. To make this precise, we will work in a basis where we separate normal operators and evanescents and write the anomalous dimension matrix as follows:
\beq
\Gamma(g) =\begin{pmatrix}\gamma_{11} & \gamma_{12} \\ \gamma_{21} & \gamma_{22}\end{pmatrix} 
\eeq
The blocks $\gamma_{11},$ $\gamma_{12}$ and $\gamma_{22}$ are non-zero at order $g$, whereas we explained above that the order $g$ contribution to $\gamma_{21}$ vanishes. There is however no selection rule that forbids contributions of order $g^2$ and higher to $\gamma_{21}$. This means that starting at two loops, scaling operators in the Wilson-Fisher theory are generically a mixture of normal operators and evanescents in the 4d free theory.

 In order to distinguish between normal operators and evanescents at higher loops, we can apply a field redefinition to the evanescents. This means that we are modifying the renormalization scheme. The mixing matrix in this new scheme~\cite{Dugan:1990df} is schematically given by
  \beq
 Z = \unit + g \left( Z_{1,0} + \frac{1}{\eps} Z_{1,1} \right) + g^2 \left( Z_{2,0} + \frac{1}{\eps} Z_{2,1} + \frac{1}{\eps^2} Z_{2,2} \right) +  O(g^3)\,.
 \eeq
 The $1/\eps^n$ counterterms cancel UV divergences as before, whereas the finite counterterms $Z_{n,0}$ are added to ensure that $\gamma_{21}$ vanishes order by order in $g$. At this stage, the matrix $\Gamma$ is block triangular, so it can be diagonalized separately for normal operators and evanescents. However, the matrix $\gamma_{11}$ now gets an explicit contribution at order $g^2$ from the evanescents, via the finite counterterms $Z_{1,0}$. An example of this phenomenon applied to four-fermion operators is described in Refs.~\cite{Buras:1989xd,Dugan:1990df}. It would be interesting to exhibit such two-loop contributions coming from evanescent operators in the Wilson-Fisher theory as well. We stress that in the scalar sector of WF, such contributions can appear only for operators of classical dimension $\DD \geq 15$. In particular, they will not affect the dimensions of $\phi$, $\phi^2$ and $\phi^4$ at any order in $\eps$. 

However, since high dimension operators appear in the OPE of low-dimension ones, the four point functions of low-dimension physical operators will be affected. In appendix~\ref{app:implications} we discuss these four-point functions, and in particular how the effects of unitary violation could in principle be observed in them.

\section{Final comments}
\label{sec:finalcomments}

\subsection{Some open problems}

It would be worthwhile to extend our analysis to other theories, and to more systematically enumerate all the causes of unitarity violation in non-integer dimensions. In particular, we would like to know if non-unitarity is always related to evanescent operators or whether other mechanisms exist that make the continuation of a theory to non-integer $d$ violate unitarity. It would also be interesting to understand the implications of non-unitarity for other observables in the theory, in particular recent work~\cite{Giombi:2014xxa,Fei:2014yja,Giombi:2015haa,Fei:2015kta,Fei:2015oha} on the extension of the $a$ central charge and the free energy $F$ to theories in non-integer dimensions.

\subsection{Spectrum continuity}
\label{sec:synth}

The standard lore says that the Wilson-Fisher fixed point in $d$ dimensions provide an `analytic continuation' which interpolates between the free theory in 4d and the Ising model critical point in 3d and 2d.\footnote{Or even down to $d=1$, see note \ref{note:Miguel}.} The nature of this analytic continuation is still not fully understood. We will now describe how one might think about it, and how the results of this paper fit into the picture.
 
The WF fixed points in $d=2,3,4$ are bona fide unitary CFTs, and are nonperturbatively defined.\footnote{For $d=3$ this has not yet been proved,
  although the evidence, recently reviewed in section 4 of \cite{Paulos:2015jfa}, is overwhelming.}
The interpolating fixed points in non-integer dimensions were initially defined only perturbatively, by analytically continuing Feynman diagrams. It's not a priori clear if they make sense nonperturbatively. However in much of the literature it is tacitly assumed that they do.

For example it is assumed that the dimensions of the most important scalar operators $\phi$, $\phi^2$ are analytic functions of $d$ which reduce to the free theory values in $d=4$ and to the dimensions of the leading $\bZ_2$-odd and even scalar primaries $\sigma$, $\vareps$ in 3d and 2d. 

\begin{figure}[htbp]
\begin{center}
\includegraphics[scale=1.0]{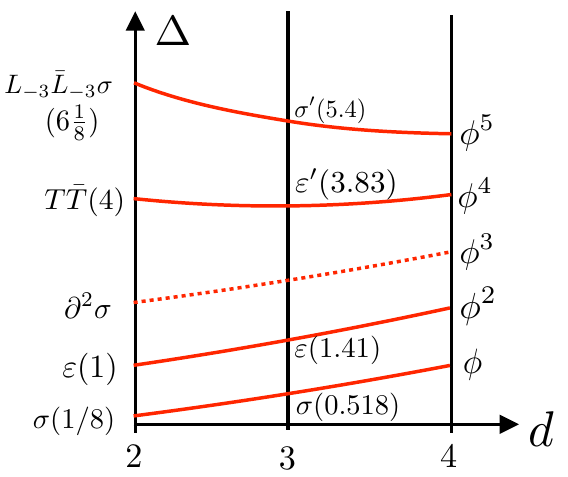}
\end{center}
\caption{Expected interpolating behavior for the low-dimension scalar operators. The exact 2d dimensions and the approximate 3d dimensions are given in parentheses. The operator $\phi^3$ is a primary in 4d but is the descendant $\del^2\sigma$ below 4d. Other descendant states are not shown.}
\label{fig:dims}
\end{figure}

Furthermore, it seems reasonable to assume that this is true not just for $\phi$, $\phi^2$ but that the whole CFT spectrum is well defined for non-integer $d$ and interpolates continuously the spectra for $d=2,3,4$ (``spectrum continuity"). E.g.~for the low dimension scalar operators we expect to see curves as in figure \ref{fig:dims}.

The spectrum continuity hypothesis seems reasonable both for primaries and for their descendants, i.e.~if a primary interpolates then its whole conformal multiplet can be assumed to interpolate. Notice that the ``number of states" in a multiplet, to the extent this concept makes sense in non-integer dimensions, has to change continuously with $d$. E.g.~if $\calO$ is a scalar primary than on the first descendant level $\del_\mu\calO$ we have $d$ states.\footnote{Conceptually speaking evanescent operators naturally emerge from the framework discussed e.g. in \cite{Collins:1984xc}, where a vector like $x^\mu$ has infinitely many components in generic $d$, but in integer $d$ all but the first $d$ components decouple. For our purposes it has not been necessary to adopt this viewpoint.} 

There are some states which are primaries in 4d but become descendants below 4d. One such state is $\phi^3$ which becomes a descendant of $\phi$ below 4d, see fig.~\ref{fig:dims}. In perturbation theory this is a consequence of the equation of motion, but more abstractly it can be seen as the phenomenon of conformal multiplet recombination, which takes place whenever a field which saturates the unitarity bound in 4d moves above the bound below 4d. The same happens for the spin-$l$ currents $\phi\overset{\leftrightarrow}{\del}{}^l\phi-\text{traces}$, $l=4,6,\ldots$, which are conserved in 4d but not in $d<4$. In $d=2$ one acquires again infinitely many conserved currents as a consequence of the Virasoro algebra. How this algebra gets broken in $d=2+\eps$ is also an interesting open question.

The results in this paper add the following ingredients to this story. First of all we have proven the existence of evanescent operators. Pictorially speaking, these appear in figure \ref{fig:dims} as new `evanescent' lines that correspond to additional states beyond those in the physical theories in integer dimensions. We don't see any consistent way which would allow one to exclude these states for non-integer $d$. Secondly, the anomalous dimensions are no longer guaranteed to be real; in figure \ref{fig:dims} there will be scaling dimensions of high-dimension evanescent operators with a non-zero imaginary part. These operators have vanishing norms in integer dimensions, so then (and only then) we can consistently remove these states and recover a unitary theory. The hypothesis of spectrum continuity gets modified accordingly: the most natural assumption is that spectrum continuity is still true in the enlarged Hilbert space which includes the evanescent operators with their potentially complex dimensions.

\subsection{Numerical bootstrap in non-integer dimensions}
\label{subsec:bootstrap}

Let us finally discuss the effects of non-unitarity for the analysis in \cite{El-Showk:2013nia} of the Wilson-Fisher fixed point using conformal bootstrap methods \cite{Rattazzi:2008pe} in $2 < d < 4$. Such an analysis requires only a definition of the conformal blocks in non-integer $d$, which is relatively straightforward: one defines the blocks as a solution of the conformal Casimir equation \cite{Dolan:2003hv}, where $d$ enters simply as a parameter. In  \cite{Rychkov:2009ij,ElShowk:2012ht,El-Showk:2014dwa} it was found that scaling dimension bounds in $d=2$ and $d=3$ exhibit kinks that essentially coincide with the location of the Ising CFT. In \cite{El-Showk:2013nia} it was shown that these kinks continue to exist in non-integer $d$ and that their location agrees with resummed perturbative results obtained from the $4-\epsilon$ expansions. 

Ref.~\cite{El-Showk:2013nia} assumed that the WF theories in $4-\epsilon$ dimensions are unitary. This assumption was crucial for the bootstrap analysis, as it implied that all squared OPE coefficient were positive, and all scaling dimensions real and consistent with unitarity bounds. We have now shown that these assumptions are in fact untrue. In spite of this fact, the results of \cite{El-Showk:2013nia} produced entirely reasonable-looking results for the dimensions of $\phi$ and $\phi^2$. This may seem at first paradoxical. 
The explanation of the paradox, already given in \cite{tcsa}, must have to do with the fact that the unitarity violation effects occur only at relatively high dimension. In a four-point function the high-dimension operators decouple exponentially fast \cite{Pappadopulo:2012jk} and consequently the numerical bootstrap methods are rather insensitive to their behavior. Also, there are many positive-norm operators which live at about the same dimension as the negative-norm ones, which leads to further suppression of unitarity-violating effects (see also appendix \ref{app:implications}).

There are other cases when the numerical bootstrap applied to non-unitary theories gave reasonable results, probably for the same reason of the mildness of the unitarity violating effects (e.g.~theories with four supercharges in non-integer dimensions \cite{Bobev:2015jxa,Chester:2015lej}, and the $O(N)$ model at $d=5.95$ \cite{Chester:2014gqa}). Surpisingly, even the 3d $O(N)$ model with non-integer $0<N<2$ seems to be in this class \cite{Shimada:2015gda}.

In other non-unitary theories the unitarity violation strength is order one and the numerical bootstrap assuming positive OPE coefficients is inadequate. This is what may have been observed for $1<d<2$ in~\cite{Golden:2014oqa}. The severe truncation methods of Gliozzi and collaborators~\cite{Gliozzi:2013ysa,Gliozzi:2014jsa,Gliozzi:2015qsa} are currently the only ones applicable in such situations.

\subsection*{Acknowledgements}
We thank Jesper Jacobsen, Miguel Paulos and Hirohiko Shimada for the useful discussions. S.R.'s research was partly supported by the National Centre of Competence in Research SwissMAP funded by the Swiss National Science Foundation.

\appendix

\section{Small block selection rule}
\label{sec:selecRule}

Let us recall the small block selection rule stated in section~\ref{sec:WFopes}: suppose that scalars $\calA$ and $\calB$ have the same 4d dimension, and that $\phi^4(x)\times \calA(0)\supset \calB(0)$.
Then $\calA$ and $\calB$ have the same number of $\phi$'s. 

We first observed this rule empirically, and then found the following proof. The same result was obtained by different means in Ref.~\cite{Kehrein:1992fn}.

\emph{Proof.} The OPE is obtained by Wick-contracting $k$ out of four $\phi$'s inside $\phi^4(x)$ with $\phi$'s inside $\calA$, expanding the remaining part of $\phi^4(x)$ in Taylor series around zero, and contracting indices to get a scalar. The number of $\phi$'s inside $\calB$ will be the same if $k=2$.

We thus have to show that any other $k$ cannot give a scalar $\calB$ of the same 4d dimension as $\calA$. 
Notice that every Wick contraction strikes out fields out of $\calA$, lowering the remaining dimension by the number of contracted fields plus the number of derivatives they carried. Once Wick contractions are finished, the dimension is raised by multiplying with what remains of $\phi^4(x)$ and by all derivatives we put on it when Taylor-expanding. The challenge is to show that these two procedures cannot compensate each other. 

The simplest cases are then $k=0$ and $k=4$. For $k=0$ the dimension is not lowered as there are no Wick contractions, while in the second step it's raised by at least four units. Analogously, for $k=4$ the dimension is only lowered by Wick contractions, by at least four units, but cannot be subsequently raised since nothing is left of $\phi^4(x)$. 

The cases $k=1$ and $k=3$ require more work. We start with $k=1$. Let us write $\calA$ in the schematic form $\calA' \del^l\phi_1$, where $\phi_1$ is the $\phi$ inside $\calA$ which gets Wick-contracted, and the notation shows that it carries $l$ derivatives. When these derivatives fall on the propagator $\langle \phi(x)\phi_1(0)\rangle$, they produce rank $l$ traceless symmetric tensor made out of $x$. To generate a scalar operator, the $\phi^3(x)$ which remained after the Wick contraction must be Taylor-expanded to the same order $l$, or higher. The generated $\calB$ state is of the schematic form $(\del^2)^n \del^l(\phi^3)\calA'$. Its 4d dimension is thus larger than that of $\calA$ by at least two units. The crux of this argument is that the extra dimension lost in the first step due to derivatives carried by $\phi_1$ is necessarily recovered in the second step if we want to get a scalar state.

Finally let's consider $k=3$. Write $\calA$ in the form $\calA'\, \del^{l}(\phi_1 \phi_2 \phi_3)$, where $\phi_i$ get Wick-contracted. The notation means that $\phi_i$'s carry $l$ derivatives in total, but they can be arbitrarily distributed among these fields. Let $l'\le l$ denote the number of these derivatives which are contracted with $\phi$'s inside $\calA'$. The rest of them are contracted among themselves. After Wick contractions, we are left with $\phi(x)\calA'(0)$ where $\calA'$ has $l'$ open indices. To get a scalar $\calB$, we must Taylor-expand $\phi(x)$ to order exactly $l'$. We thus get $\calB$ of the form $\del^{l'}\phi\, \calA'$. The resulting 4d dimension is lower than that of $\calA$ by at least 2 units. The equation of motion of $\phi$ was important here, preventing us from considering the states of the form $(\del^2)^n\del^{l'}\phi\, \calA'$.

\section{Non-unitarity and four-point functions}
\label{app:implications}
The non-unitarity of the theory in non-integer $d$ is intrinsically linked to operators of high dimension. However, because of the OPE, we should be able to observe the non-unitarity also by considering higher-point correlation functions of `simple' low-dimension operators like $\phi^2(x)$ and $\phi^3(x)$. In this appendix we briefly discuss the consequences of non-unitarity in the specific example of a four-point function of identical Hermitian scalar operators $\OO(x)$.

Conformal invariance dictates that
\be
\expec{\OO (x_1) \ldots \OO (x_4)} = \frac{1}{x_{12}^{2 \D_{\OO}} x_{34}^{2 \D_\OO}} G(u,v)
\ee
with $u$ and $v$ the familiar cross ratios. The appearance of a primary operator $\mathcal K(x)$ of dimension $\D$ and spin $\ell$ in the $\OO(x_1) \OO(x_2)$ OPE leads to a term in $G(u,v)$ of the form
\be
G(u,v) = \ldots + a u^{(\D-\ell)/2} (1-v)^\ell + \ldots
\ee
which is the leading term corresponding to this multiplet as $u \to 0$ and (subsequently) $v \to 1$. If $\mathcal K(x)$ is Hermitian with a positive/negative norm then $a$ is positive/negative. If $\mathcal K(x)$ is not Hermitian then its conjugate also appears in the OPE, with conjugate coefficient, leading to
\be
\begin{split}
G(u,v) &= \ldots + (a u^{(\D- \ell)/2} + \bar a u^{(\bar \Delta -\ell)/2}) (1-v)^\ell + \ldots\\
&=\ldots + 2 |a| u^{(\D_1 - \ell)/2}\cos(\D_2 \log(u)/2  + \theta) (1-v)^\ell + \ldots
\end{split}
\ee
with $\D = \D_1 + i \D_2$ and $a = |a| \exp( i \theta)$. Clearly, both negative norm states as well as complex dimensions lead to very distinct behavior in the OPE limit.

The non-unitarity might be harder to detect in perturbation theory. In that case both the coefficient $a$ and the dimension $\Delta$ will have perturbative expansions. We write
\be
a = a^0 + \delta a\,, \qquad \Delta = \Delta^0 + \gamma\,.
\ee
Unfortunately there are in general other primary operators with the same spin and $\Delta^0$ that appear in the conformal block decomposition. Let us assume that there are $K$ of them, and denote them with a subscript $i$. We then find
\be
G(u,v) = \ldots + u^{(\Delta^0 - \ell)/2} (1-v)^\ell \sum_{n=0}^\infty X_n \log(u)^n + \ldots
\ee
with
\be
\label{eq:Xn}
X_n = \sum_{i=1}^K (a^0_i + \delta a_i) (\gamma_i)^n\,.
\ee
In the epsilon expansion the $\gamma_i$ are $O(\epsilon)$ and therefore the coefficients $X_n$ are $O(\epsilon^{n})$. On the right-hand side of \eqref{eq:Xn} we have $2 K$ independent parameters, so in general the $X_{n \geq 2K}$ can be expressed in terms of the $X_{0 \leq n < 2K}$. Conversely, in order to resolve even the one-loop corrections to $\delta a_i$ and $\gamma_i$ we need to know the leading order behavior of $X_{0 \leq n < 2K}$, which requires knowledge of the correction at order $\epsilon^{2K-1}$. We conclude that the non-unitarity might show up only at relatively high loop order in the four-point functions.

\small
\parskip=-10pt
\bibliography{biblio}

\providecommand{\href}[2]{#2}\begingroup\raggedright\begin{thebibliography}{10}

\bibitem{tcsa}
M.~Hogervorst, S.~Rychkov, and B.~C. van Rees, ``{Truncated conformal space
  approach in $d$ dimensions: A cheap alternative to lattice field theory?},''
  \href{http://dx.doi.org/10.1103/PhysRevD.91.025005}{{\em Phys. Rev.}
  {\bfseries D91} (2015) 025005},
\href{http://arxiv.org/abs/1409.1581}{{\ttfamily arXiv:1409.1581 [hep-th]}}.

\bibitem{Wilson:1971dc}
K.~G. Wilson and M.~E. Fisher, ``{Critical Exponents in 3.99 Dimensions},''
\href{http://dx.doi.org/10.1103/PhysRevLett.28.240}{{\em Phys.Rev.Lett.}
  {\bfseries 28} (1972) 240--243}.

\bibitem{Golden:2014oqa}
J.~Golden and M.~F. Paulos, ``{No unitary bootstrap for the fractal Ising
  model},'' \href{http://dx.doi.org/10.1007/JHEP03(2015)167}{{\em JHEP}
  {\bfseries 03} (2015) 167},
\href{http://arxiv.org/abs/1411.7932}{{\ttfamily arXiv:1411.7932 [hep-th]}}.

\bibitem{Collins:1984xc}
J.~C. Collins, {\em {Renormalization}}.
\newblock Cambridge University Press,
1984.
\newblock

\bibitem{Dugan:1990df}
M.~J. Dugan and B.~Grinstein, ``{On the vanishing of evanescent operators},''
\href{http://dx.doi.org/10.1016/0370-2693(91)90680-O}{{\em Phys. Lett.}
  {\bfseries B256} (1991) 239--244}.

\bibitem{Gracey:2008mf}
J.~A. Gracey, ``{Four loop MS-bar mass anomalous dimension in the Gross-Neveu
  model},'' \href{http://dx.doi.org/10.1016/j.nuclphysb.2008.04.002}{{\em Nucl.
  Phys.} {\bfseries B802} (2008) 330--350},
\href{http://arxiv.org/abs/0804.1241}{{\ttfamily arXiv:0804.1241 [hep-th]}}.

\bibitem{Bos:1987fb}
M.~Bos, ``{An Example of Dimensional Regularization With Antisymmetric
  Tensors},''
\href{http://dx.doi.org/10.1016/0003-4916(88)90164-9}{{\em Annals Phys.}
  {\bfseries 181} (1988) 177}.

\bibitem{Buras:1998raa}
A.~J. Buras, ``{Weak Hamiltonian, CP violation and rare decays},'' in {\em
  {Probing the standard model of particle interactions. Proceedings, Summer
  School in Theoretical Physics, NATO Advanced Study Institute, 68th session,
  Les Houches, France, July 28-September 5, 1997. Pt. 1, 2}}, pp.~281--539.
\newblock 1998.
\newblock
\href{http://arxiv.org/abs/hep-ph/9806471}{{\ttfamily arXiv:hep-ph/9806471
  [hep-ph]}}.
\newblock

\bibitem{Kehrein:1992fn}
S.~Kehrein, F.~Wegner, and Y.~Pismak, ``{Conformal symmetry and the spectrum of
  anomalous dimensions in the $N$-vector model in $4-\epsilon$ dimensions},''
\href{http://dx.doi.org/10.1016/0550-3213(93)90124-8}{{\em Nucl.Phys.}
  {\bfseries B402} (1993) 669--692}.

\bibitem{Kehrein:1994ff}
S.~K. Kehrein and F.~Wegner, ``{The Structure of the spectrum of anomalous
  dimensions in the $N$-vector model in $4-\epsilon$ dimensions},''
  \href{http://dx.doi.org/10.1016/0550-3213(94)90406-5}{{\em Nucl.Phys.}
  {\bfseries B424} (1994) 521--546},
\href{http://arxiv.org/abs/hep-th/9405123}{{\ttfamily arXiv:hep-th/9405123
  [hep-th]}}.

\bibitem{Kehrein:1995ia}
S.~K. Kehrein, ``{The spectrum of critical exponents in $(\vec\phi^2)^2$-theory
  in $d= 4-\epsilon$ dimensions: Resolution of degeneracies and hierarchical
  structures},'' \href{http://dx.doi.org/10.1016/0550-3213(95)00375-3}{{\em
  Nucl.Phys.} {\bfseries B453} (1995) 777--806},
\href{http://arxiv.org/abs/hep-th/9507044}{{\ttfamily arXiv:hep-th/9507044
  [hep-th]}}.

\bibitem{Gaiotto:2013nva}
D.~Gaiotto, D.~Mazac, and M.~F. Paulos, ``{Bootstrapping the 3d Ising twist
  defect},'' \href{http://dx.doi.org/10.1007/JHEP03(2014)100}{{\em JHEP}
  {\bfseries 03} (2014) 100},
\href{http://arxiv.org/abs/1310.5078}{{\ttfamily arXiv:1310.5078 [hep-th]}}.

\bibitem{Chester:2015wao}
S.~M. Chester, M.~Mezei, S.~S. Pufu, and I.~Yaakov, ``{Monopole operators from
  the $4-\epsilon$ expansion},''
\href{http://arxiv.org/abs/1511.07108}{{\ttfamily arXiv:1511.07108 [hep-th]}}.

\bibitem{Wilson:1972cf}
K.~G. Wilson, ``{Quantum field theory models in less than four dimensions},''
\href{http://dx.doi.org/10.1103/PhysRevD.7.2911}{{\em Phys.Rev.} {\bfseries D7}
  (1973) 2911--2926}.

\bibitem{feigin1988}
B.~L. Feigin, ``{The Lie algebras $\mathfrak{gl}(\lambda)$ and cohomologies of
  Lie algebras of differential operators},''
  \href{http://dx.doi.org/10.1070/RM1988v043n02ABEH001720}{{\em Russian
  Mathematical Surveys} {\bfseries 43} no.~2, (1988) 169}.

\bibitem{Cardy:1991kr}
J.~L. Cardy, ``{Operator content and modular properties of higher dimensional
  conformal field theories},''
\href{http://dx.doi.org/10.1016/0550-3213(91)90024-R}{{\em Nucl.Phys.}
  {\bfseries B366} (1991) 403--419}.

\bibitem{francesco_relations_1987}
P.~d. Francesco, H.~Saleur, and J.~B. Zuber, ``Relations between the {Coulomb}
  gas picture and conformal invariance of two-dimensional critical models,''
  \href{http://dx.doi.org/10.1007/BF01009954}{{\em J. Stat. Phys.} {\bfseries
  49} no.~1-2, (1987) 57--79}.

\bibitem{Cardy:2006fg}
J.~L. Cardy, ``{The $O(n)$ model on the annulus},''
  \href{http://dx.doi.org/10.1007/s10955-006-9186-8}{{\em J. Statist. Phys.}
  {\bfseries 125} (2006) 1},
\href{http://arxiv.org/abs/math-ph/0604043}{{\ttfamily arXiv:math-ph/0604043
  [math-ph]}}.

\bibitem{Hogervorst:2015tka}
M.~Hogervorst, {\em {Two studies on conformal and strongly coupled quantum
  field theories in $d>2$ dimensions}}.
\newblock PhD thesis, {\'{E}cole normale sup\'{e}rieure}, 2015.
\newblock
\url{https://tel.archives-ouvertes.fr/tel-01171444}.
\newblock

\bibitem{Luscher:1974ez}
M.~Luscher and G.~Mack, ``{Global Conformal Invariance in Quantum Field
  Theory},''
\href{http://dx.doi.org/10.1007/BF01608988}{{\em Commun. Math. Phys.}
  {\bfseries 41} (1975) 203--234}.

\bibitem{SRnotes}
S.~Rychkov, ``{EPFL Lectures on Conformal Field Theory in $D \ge 3$
  Dimensions},''
\href{http://arxiv.org/abs/1601.05000}{{\ttfamily arXiv:1601.05000 [hep-th]}}.

\bibitem{Maldacena:2011jn}
J.~Maldacena and A.~Zhiboedov, ``{Constraining Conformal Field Theories with A
  Higher Spin Symmetry},''
  \href{http://dx.doi.org/10.1088/1751-8113/46/21/214011}{{\em J. Phys.}
  {\bfseries A46} (2013) 214011},
\href{http://arxiv.org/abs/1112.1016}{{\ttfamily arXiv:1112.1016 [hep-th]}}.

\bibitem{ElShowk:2011ag}
S.~El-Showk and K.~Papadodimas, ``{Emergent Spacetime and Holographic CFTs},''
  \href{http://dx.doi.org/10.1007/JHEP10(2012)106}{{\em JHEP} {\bfseries 1210}
  (2012) 106},
\href{http://arxiv.org/abs/1101.4163}{{\ttfamily arXiv:1101.4163 [hep-th]}}.

\bibitem{Pappadopulo:2012jk}
D.~Pappadopulo, S.~Rychkov, J.~Espin, and R.~Rattazzi, ``{OPE Convergence in
  Conformal Field Theory},''
  \href{http://dx.doi.org/10.1103/PhysRevD.86.105043}{{\em Phys. Rev.}
  {\bfseries D86} (2012) 105043},
\href{http://arxiv.org/abs/1208.6449}{{\ttfamily arXiv:1208.6449 [hep-th]}}.

\bibitem{Larin:1993tq}
S.~A. Larin, ``{The Renormalization of the axial anomaly in dimensional
  regularization},'' \href{http://dx.doi.org/10.1016/0370-2693(93)90053-K}{{\em
  Phys. Lett.} {\bfseries B303} (1993) 113--118},
\href{http://arxiv.org/abs/hep-ph/9302240}{{\ttfamily arXiv:hep-ph/9302240
  [hep-ph]}}.

\bibitem{Gracey:2000am}
J.~A. Gracey, ``{Three loop MS-bar tensor current anomalous dimension in
  QCD},'' \href{http://dx.doi.org/10.1016/S0370-2693(00)00859-5}{{\em Phys.
  Lett.} {\bfseries B488} (2000) 175--181},
\href{http://arxiv.org/abs/hep-ph/0007171}{{\ttfamily arXiv:hep-ph/0007171
  [hep-ph]}}.

\bibitem{Kleinert}
H.~Kleinert and V.~Schulte-Frohlinde, {\em Critical Properties of
  $\phi^4$-Theories}.
\newblock World Scientific, 2001.

\bibitem{Vasiliev}
A.~Vasil'ev, {\em The Field Theoretic Renormalization Group in Critical
  Behaviour Theory and Stochastic Dynamics}.
\newblock CRC Press, 2004.

\bibitem{Cardy}
J.~L. Cardy, {\em {Scaling and Renormalization in Statistical Physics}}.
\newblock Cambridge University Press,
1996.
\newblock

\bibitem{Rychkov:2015naa}
S.~Rychkov and Z.~M. Tan, ``{The $\epsilon$-expansion from conformal field
  theory},'' \href{http://dx.doi.org/10.1088/1751-8113/48/29/29FT01}{{\em J.
  Phys.} {\bfseries A48} no.~29, (2015) 29FT01},
\href{http://arxiv.org/abs/1505.00963}{{\ttfamily arXiv:1505.00963 [hep-th]}}.

\bibitem{Basu:2015gpa}
P.~Basu and C.~Krishnan, ``{$\epsilon$-expansions near three dimensions from
  conformal field theory},''
  \href{http://dx.doi.org/10.1007/JHEP11(2015)040}{{\em JHEP} {\bfseries 11}
  (2015) 040},
\href{http://arxiv.org/abs/1506.06616}{{\ttfamily arXiv:1506.06616 [hep-th]}}.

\bibitem{Raju:2015fza}
A.~Raju, ``{$\epsilon$-Expansion in the Gross-Neveu CFT},''
\href{http://arxiv.org/abs/1510.05287}{{\ttfamily arXiv:1510.05287 [hep-th]}}.

\bibitem{Ghosh:2015opa}
S.~Ghosh, R.~K. Gupta, K.~Jaswin, and A.~A. Nizami, ``{$\epsilon$-Expansion in
  the Gross-Neveu Model from Conformal Field Theory},''
\href{http://arxiv.org/abs/1510.04887}{{\ttfamily arXiv:1510.04887 [hep-th]}}.

\bibitem{Sen:2015doa}
K.~Sen and A.~Sinha, ``{On critical exponents without Feynman diagrams},''
\href{http://arxiv.org/abs/1510.07770}{{\ttfamily arXiv:1510.07770 [hep-th]}}.

\bibitem{Mack:1975je}
G.~Mack, ``{All Unitary Ray Representations of the Conformal Group SU(2,2) with
  Positive Energy},''
\href{http://dx.doi.org/10.1007/BF01613145}{{\em Commun.Math.Phys.} {\bfseries
  55} (1977) 1}.

\bibitem{Minwalla:1997ka}
S.~Minwalla, ``{Restrictions imposed by superconformal invariance on quantum
  field theories},'' {\em Adv.Theor.Math.Phys.} {\bfseries 2} (1998) 781--846,
\href{http://arxiv.org/abs/hep-th/9712074}{{\ttfamily arXiv:hep-th/9712074
  [hep-th]}}.

\bibitem{Buras:1989xd}
A.~J. Buras and P.~H. Weisz, ``{QCD Nonleading Corrections to Weak Decays in
  Dimensional Regularization and 't Hooft-Veltman Schemes},''
\href{http://dx.doi.org/10.1016/0550-3213(90)90223-Z}{{\em Nucl. Phys.}
  {\bfseries B333} (1990) 66}.

\bibitem{Giombi:2014xxa}
S.~Giombi and I.~R. Klebanov, ``{Interpolating between $a$ and $F$},''
  \href{http://dx.doi.org/10.1007/JHEP03(2015)117}{{\em JHEP} {\bfseries 03}
  (2015) 117},
\href{http://arxiv.org/abs/1409.1937}{{\ttfamily arXiv:1409.1937 [hep-th]}}.

\bibitem{Fei:2014yja}
L.~Fei, S.~Giombi, and I.~R. Klebanov, ``{Critical $O(N)$ models in $6 -
  \epsilon$ dimensions},''
  \href{http://dx.doi.org/10.1103/PhysRevD.90.025018}{{\em Phys.Rev.}
  {\bfseries D90} no.~2, (2014) 025018},
\href{http://arxiv.org/abs/1404.1094}{{\ttfamily arXiv:1404.1094 [hep-th]}}.

\bibitem{Giombi:2015haa}
S.~Giombi, I.~R. Klebanov, and G.~Tarnopolsky, ``{Conformal QED$_d$,
  $F$-Theorem and the $\epsilon$ Expansion},''
  \href{http://dx.doi.org/10.1088/1751-8113/49/13/135403}{{\em J. Phys.}
  {\bfseries A49} no.~13, (2016) 135403},
\href{http://arxiv.org/abs/1508.06354}{{\ttfamily arXiv:1508.06354 [hep-th]}}.

\bibitem{Fei:2015kta}
L.~Fei, S.~Giombi, I.~R. Klebanov, and G.~Tarnopolsky, ``{Critical $Sp(N)$
  models in $6 - \epsilon$ dimensions and higher spin dS/CFT},''
  \href{http://dx.doi.org/10.1007/JHEP09(2015)076}{{\em JHEP} {\bfseries 09}
  (2015) 076},
\href{http://arxiv.org/abs/1502.07271}{{\ttfamily arXiv:1502.07271 [hep-th]}}.

\bibitem{Fei:2015oha}
L.~Fei, S.~Giombi, I.~R. Klebanov, and G.~Tarnopolsky, ``{Generalized
  $F$-Theorem and the $\epsilon$ Expansion},''
  \href{http://dx.doi.org/10.1007/JHEP12(2015)155}{{\em JHEP} {\bfseries 12}
  (2015) 155},
\href{http://arxiv.org/abs/1507.01960}{{\ttfamily arXiv:1507.01960 [hep-th]}}.

\bibitem{Paulos:2015jfa}
M.~F. Paulos, S.~Rychkov, B.~C. van Rees, and B.~Zan, ``{Conformal Invariance
  in the Long-Range Ising Model},''
  \href{http://dx.doi.org/10.1016/j.nuclphysb.2015.10.018}{{\em Nucl. Phys.}
  {\bfseries B902} (2016) 246--291},
\href{http://arxiv.org/abs/1509.00008}{{\ttfamily arXiv:1509.00008 [hep-th]}}.

\bibitem{El-Showk:2013nia}
S.~El-Showk, M.~Paulos, D.~Poland, S.~Rychkov, D.~Simmons-Duffin, and A.~Vichi,
  ``{Conformal Field Theories in Fractional Dimensions},''
  \href{http://dx.doi.org/10.1103/PhysRevLett.112.141601}{{\em Phys.Rev.Lett.}
  {\bfseries 112} (2014) 141601},
\href{http://arxiv.org/abs/1309.5089}{{\ttfamily arXiv:1309.5089 [hep-th]}}.

\bibitem{Rattazzi:2008pe}
R.~Rattazzi, V.~S. Rychkov, E.~Tonni, and A.~Vichi, ``{Bounding scalar operator
  dimensions in 4D CFT},''
  \href{http://dx.doi.org/10.1088/1126-6708/2008/12/031}{{\em JHEP} {\bfseries
  12} (2008) 031},
\href{http://arxiv.org/abs/0807.0004}{{\ttfamily arXiv:0807.0004 [hep-th]}}.

\bibitem{Dolan:2003hv}
F.~A. Dolan and H.~Osborn, ``{Conformal partial waves and the operator product
  expansion},'' \href{http://dx.doi.org/10.1016/j.nuclphysb.2003.11.016}{{\em
  Nucl. Phys.} {\bfseries B678} (2004) 491--507},
\href{http://arxiv.org/abs/hep-th/0309180}{{\ttfamily arXiv:hep-th/0309180
  [hep-th]}}.

\bibitem{Rychkov:2009ij}
V.~S. Rychkov and A.~Vichi, ``{Universal Constraints on Conformal Operator
  Dimensions},'' \href{http://dx.doi.org/10.1103/PhysRevD.80.045006}{{\em Phys.
  Rev.} {\bfseries D80} (2009) 045006},
\href{http://arxiv.org/abs/0905.2211}{{\ttfamily arXiv:0905.2211 [hep-th]}}.

\bibitem{ElShowk:2012ht}
S.~El-Showk, M.~F. Paulos, D.~Poland, S.~Rychkov, D.~Simmons-Duffin, and
  A.~Vichi, ``{Solving the 3D Ising Model with the Conformal Bootstrap},''
  \href{http://dx.doi.org/10.1103/PhysRevD.86.025022}{{\em Phys. Rev.}
  {\bfseries D86} (2012) 025022},
\href{http://arxiv.org/abs/1203.6064}{{\ttfamily arXiv:1203.6064 [hep-th]}}.

\bibitem{El-Showk:2014dwa}
S.~El-Showk, M.~F. Paulos, D.~Poland, S.~Rychkov, D.~Simmons-Duffin, and
  A.~Vichi, ``{Solving the 3d Ising Model with the Conformal Bootstrap II.
  c-Minimization and Precise Critical Exponents},''
  \href{http://dx.doi.org/10.1007/s10955-014-1042-7}{{\em J. Stat. Phys.}
  {\bfseries 157} (2014) 869},
\href{http://arxiv.org/abs/1403.4545}{{\ttfamily arXiv:1403.4545 [hep-th]}}.

\bibitem{Bobev:2015jxa}
N.~Bobev, S.~El-Showk, D.~Mazac, and M.~F. Paulos, ``{Bootstrapping SCFTs with
  Four Supercharges},'' \href{http://dx.doi.org/10.1007/JHEP08(2015)142}{{\em
  JHEP} {\bfseries 08} (2015) 142},
\href{http://arxiv.org/abs/1503.02081}{{\ttfamily arXiv:1503.02081 [hep-th]}}.

\bibitem{Chester:2015lej}
S.~M. Chester, L.~V. Iliesiu, S.~S. Pufu, and R.~Yacoby, ``{Bootstrapping
  $O(N)$ Vector Models with Four Supercharges in $3 \leq d \leq4$},''
\href{http://arxiv.org/abs/1511.07552}{{\ttfamily arXiv:1511.07552 [hep-th]}}.

\bibitem{Chester:2014gqa}
S.~M. Chester, S.~S. Pufu, and R.~Yacoby, ``{Bootstrapping $O(N)$ vector models
  in 4 $< d <$ 6},'' \href{http://dx.doi.org/10.1103/PhysRevD.91.086014}{{\em
  Phys. Rev.} {\bfseries D91} no.~8, (2015) 086014},
\href{http://arxiv.org/abs/1412.7746}{{\ttfamily arXiv:1412.7746 [hep-th]}}.

\bibitem{Shimada:2015gda}
H.~Shimada and S.~Hikami, ``{Fractal dimensions of self-avoiding walks and
  Ising high-temperature graphs in 3D conformal bootstrap},''
\href{http://arxiv.org/abs/1509.04039}{{\ttfamily arXiv:1509.04039
  [cond-mat.stat-mech]}}.

\bibitem{Gliozzi:2013ysa}
F.~Gliozzi, ``{More constraining conformal bootstrap},''
  \href{http://dx.doi.org/10.1103/PhysRevLett.111.161602}{{\em Phys. Rev.
  Lett.} {\bfseries 111} (2013) 161602},
\href{http://arxiv.org/abs/1307.3111}{{\ttfamily arXiv:1307.3111}}.

\bibitem{Gliozzi:2014jsa}
F.~Gliozzi and A.~Rago, ``{Critical exponents of the 3d Ising and related
  models from Conformal Bootstrap},''
  \href{http://dx.doi.org/10.1007/JHEP10(2014)042}{{\em JHEP} {\bfseries 10}
  (2014) 042},
\href{http://arxiv.org/abs/1403.6003}{{\ttfamily arXiv:1403.6003 [hep-th]}}.

\bibitem{Gliozzi:2015qsa}
F.~Gliozzi, P.~Liendo, M.~Meineri, and A.~Rago, ``{Boundary and Interface CFTs
  from the Conformal Bootstrap},''
  \href{http://dx.doi.org/10.1007/JHEP05(2015)036}{{\em JHEP} {\bfseries 05}
  (2015) 036},
\href{http://arxiv.org/abs/1502.07217}{{\ttfamily arXiv:1502.07217 [hep-th]}}.

\end{thebibliography}\endgroup
\bibliographystyle{utphys.bst}
\end{document}